\documentclass[%
 aip,
jcp,%
 amsmath,amssymb,
 reprint,%
]{revtex4-1}

\usepackage{graphicx}
\usepackage{dcolumn}
\usepackage{bm}

\usepackage{xcolor}

\usepackage{amsfonts}
\usepackage{hyperref}
\usepackage[version=3]{mhchem} 
\usepackage[T1]{fontenc}
\usepackage{natbib}

\usepackage{amssymb,latexsym}
\usepackage{amsbsy} 
\usepackage{amsmath}
\usepackage{hyperref}
\usepackage{graphicx}
\usepackage{ifthen}

\newcommand \bra[1] {\langle{#1}|}
\newcommand \ket[1] {|{#1}\rangle}
\newcommand \bravac {\bra{\text{vac}}}
\newcommand \vac {\ket{\text{vac}}}

\newcommand \trace[1] {\mathrm{tr}({#1})}

\newcommand \op[1] {\hat{#1}}
\newcommand \aop[1] {\op{a}_{{#1}}}
\newcommand \aopU[1] {\op{a}^{{#1}}}
\newcommand \cop[1] {\op{a}^{\dagger}_{{#1}}}
\newcommand \copU[1] {\op{a}^{{#1}\dagger}}

\newcommand \sao[1] {
	\ifthenelse{\equal{#1}{1}}{{\alpha}}{}
	\ifthenelse{\equal{#1}{2}}{{\beta}}{}
	\ifthenelse{\equal{#1}{3}}{{\gamma}}{}
	\ifthenelse{\equal{#1}{4}}{{\epsilon}}{}
	\ifthenelse{\equal{#1}{5}}{{\zeta}}{}
	\ifthenelse{\equal{#1}{6}}{{\eta}}{}
	\ifthenelse{\equal{#1}{7}}{{\theta}}{}
	\ifthenelse{\equal{#1}{8}}{{\kappa}}{}
	\ifthenelse{\equal{#1}{9}}{{\lambda}}{}
	\ifthenelse{\equal{#1}{10}}{{\mu}}{}
	\ifthenelse{\equal{#1}{11}}{{\nu}}{}
	\ifthenelse{\equal{#1}{12}}{{\xi}}{}
	\ifthenelse{\equal{#1}{13}}{{\rho}}{}
	\ifthenelse{\equal{#1}{14}}{{\phi}}{}
	\ifthenelse{\equal{#1}{15}}{{\chi}}{}
	\ifthenelse{\equal{#1}{16}}{{\omega}}{}
}

\newcommand \GammaExp[4] {P_{{#1}{#2}} P_{{#3}{#4}} - P_{{#1}{#4}} P_{{#3}{#2}}}

\newcommand \cEcom[3] {-S_{{#3}{#1}} \cop{{#2}}}
\newcommand \aEcom[3] {S_{{#1}{#2}} \aop{{#3}}}
\newcommand \EEcom[4] {S_{{#2}{#3}} \op{E}_{{#1}{#4}} - S_{{#4}{#1}} \op{E}_{{#3}{#2}}}
\newcommand \eEcom[6] {
	S_{{#4}{#5}} \op{e}_{{#1}{#2}{#3}{#6}}
	+ S_{{#2}{#5}} \op{e}_{{#1}{#6}{#3}{#4}}
	- S_{{#6}{#1}} \op{e}_{{#5}{#2}{#3}{#4}}
	- S_{{#6}{#3}} \op{e}_{{#1}{#2}{#5}{#4}}
}

\newcommand \tDU[3] {{#1}^{\phantom{{#2}} {#3}}_{{#2}}}
\newcommand \tUD[3] {{#1}^{{#2}}_{\phantom{{#2}} {#3}}}

\newcommand \tDUUU[5] {{#1}^{\phantom{{#2}} {#3}{#4}{#5}}_{{#2}}}
\newcommand \tUDUU[5] {{#1}^{{#2} \phantom{{#3}} {#4}{#5}}_{\phantom{{#2}} {#3}}}
\newcommand \tUUDU[5] {{#1}^{{#2}{#3} \phantom{{#4}} {#5}}_{\phantom{{#2}{#3}} {#4}}}
\newcommand \tUUUD[5] {{#1}^{{#2}{#3}{#4}}_{\phantom{{#2}{#3}{#4}} {#5}}}

\newcommand \tDDUU[5] {{#1}^{\phantom{{#2}} \phantom{{#3}} {#4} {#5}}_{{#2} {#3}}}
\newcommand \tDUUD[5] {{#1}^{\phantom{{#2}} {#3} {#4}}_{{#2} \phantom{{#3}} \phantom{{#4}} {#5}}}
\newcommand \tUDUD[5] {{#1}^{{#2} \phantom{{#3}} {#4}}_{\phantom{{#2}} {#3} \phantom{{#4}} {#5}}}

\newcommand \tUUDD[5] {{#1}^{{#2} {#3}}_{\phantom{{#2}} \phantom{{#3}} {#4} {#5}}}
\newcommand \tDUDU[5] {{#1}^{\phantom{{#2}} {#3} \phantom{{#4}} {#5}}_{{#2} \phantom{{#3}} {#4}}}

\begin{document}
	
	\preprint{AIP/123-QED}
	
	\title{Excited states of molecules in strong uniform and non-uniform magnetic fields}

\date{\today}

\author{Sangita Sen}

\email{sangita.sen310187@gmail.com}
\affiliation{
	Hylleraas Centre for Quantum Molecular Sciences, Department of Chemistry, University of Oslo, P.O.~Box 1033 Blindern, N-0315 Oslo, Norway}

\author{Kai K. Lange}

\affiliation{
	Hylleraas Centre for Quantum Molecular Sciences, Department of Chemistry, University of Oslo, P.O.~Box 1033 Blindern, N-0315 Oslo, Norway}

\author{Erik I. Tellgren}

\email{erik.tellgren@kjemi.uio.no}
\affiliation{
	Hylleraas Centre for Quantum Molecular Sciences, Department of Chemistry, University of Oslo, P.O.~Box 1033 Blindern, N-0315 Oslo, Norway}

\begin{abstract}
This paper reports an implementation of Hartree-Fock linear response with complex orbitals for computing electronic spectra of molecules in a strong external magnetic fields. The implementation is completely general, allowing for spin-restricted, spin-unrestricted, and general two-component reference states. The method is applied to small molecules placed in strong uniform and non-uniform magnetic fields of astrochemical importance at the Random Phase Approximation level of theory. For uniform fields, where comparison is possible, the spectra are found to be qualitatively similar to those recently obtained with equation of motion coupled cluster theory. We also study the behaviour of spin-forbidden excitations with progressive loss of spin symmetry induced by non-uniform magnetic fields. Finally, the equivalence of length and velocity gauges for oscillator strengths when using complex orbitals is investigated and found to hold numerically.	
\end{abstract}

\maketitle

\section{Introduction}

External magnetic fields can dramatically affect the electronic structure of atoms and molecules when the field interaction strengths are comparable to the Coulomb interaction~\cite{Garstang1977, Lai2001}. This turns out to be of the order of 1~a.u $\approx$ 235~kT.
In nature, such strengths are known to exist on magnetized stellar objects such as magnetized white dwarf stars but they are two-three orders of magnitude beyond what can presently be produced in terrestrial experiments~\cite{BYKOV_PB294_574,NAKAMURA_RSI84_044702,NAKAMURA_RSI89_095106}.
The observed electronic spectra from magnetized white dwarf stars are strongly distorted by the magnetic fields making them impossible to interpret without computational support.
He, C and O have been detected so far, in addition to H~\cite{Jordan1998,Jordan2001,Liebert2003}. 
Recently, H$_2$ has been detected in non-magnetized white-dwarfs~\cite{Xu2013a}.
The possibility of small hydrocarbons cannot be ruled out either~\cite{Schmidt1995}.
The first computational efforts primarily by Ivanov and Schmelcher~\cite{Ivanov1997,Ivanov1999,Ivanov2001,Ivanov2001a} were targeted at ground and excited states of small atoms at the Hartree-Fock level. Later work focussed on few electron systems such as H$_2$, He, He$_2$, Li and Be  at the full configuration interaction (FCI) level~\cite{Lange2012,Detmer1997,Detmer1998,Becken1999,Becken2000,Becken2001,AlHujaj2004,AlHujaj2004a}. Most recently, the coupled-cluster theory (CCSD) has been used to compute ground states of atoms and molecules in strong magnetic fields~\cite{Stopkowicz2015} followed by the equation-of-motion coupled cluster treatment (EOM-CCSD) for excited states~\cite{Hampe2017}.

In this paper we present the first implementation of the linear response of the Hartree-Fock method with complex orbitals for computation of electronic spectra in an external magnetic field.
Earlier work in the non-relativistic domain has focussed on spin frustrated systems \cite{YAMAKI_IJQC80_701,Goings2015}.
The ground state is optimized in the presence of an external magnetic field and the excited states are obtained via linear response.
London atomic orbitals (LAOs) are employed to enforce gauge-origin
invariance and accelerate basis set
convergence~\cite{London1937,Hameka1958,Ditchfield1976,Helgaker1991}. With
ordinary Gaussians it becomes necessary to use very large basis sets
to approach  gauge-origin
invariance~\cite{Faglioni2004,Caputo1994,Caputo1994a,Caputo1996,Caputo1997}. An
implementation of integral evaluation for the LAOs which are
plane-wave/Gaussian hybrid functions is this
necessary~\cite{Tellgren2008,REYNOLDS_PCCP17_14280,IRONS_JCTC13_3636,SUN_JCTC15_348}. Our
implementation builds on our previous work on non-uniform magnetic
fields~\cite{TELLGREN_JCP139_164118} and General Hartree--Fock theory~\cite{Sen2018a} within the {\sc London} program~\cite{Tellgren2008,londonProgram}.
Since only the one-electron part of the Hamiltonian is modified in such a finite-field approach, no additional effort is required for extension to post-Hartree--Fock theories or for linear response, in this case.
It therefore opens up the possibility of studying non-perturbative phenomena.

Linear response provides computationally cheap access to a large number of excited states.
This is beneficial for the interpretation of complicated spectra where a large number of states are involved.
While the role of differential electron correlation between the ground and excited state is certainly important, computational results in the literature (without magnetic fields) have clearly demonstrated that linear response spectra are adequate in most cases if the ground-state is well described as in coupled cluster linear response (CC-LRT)~\cite{Watts2008,Krylov2008,Sauer2009,Koch1990c}, time dependent density functional theory (TD-DFT)~\cite{Elliott2009,Laurent2013} or multi-configurational time-dependent Hartree-Fock (MCTDHF)~\cite{Jorgensen1975}.
It has been recently shown that excitation energies from the Random Phase Approximation (RPA) correspond to an approximated EOM-CCD~\cite{Berkelbach2018}. 
For example, in this paper we have demonstrated that the evolution of the spectra of the carbon atom with changing magnetic fields is qualitatively very similar to the EOM-CCSD results by Hampe and Stopkowicz~\cite{Hampe2017}.

In addition to the possibilities for supporting spectral detection of atoms and molecules in stars, the study of excited states in strong magnetic fields is also an unexplored field as of today.
Non-perturbative transition from closed-shell para- to diamagnetism~\cite{Tellgren2009} and a new bonding mechanism~\cite{Zaucer1978,Lange2012,Tellgren2012,Kubo2007} in very strong magnetic fields have recently been computationally uncovered for ground states.
The usually more sensitive electronic structure of excited states gives rise to the possibility of discovering interesting field-induced phenomena at field strengths lower than that for ground states. 
Moreover, the response of a molecule to a magnetic field is found to increase with increase in the area of cross-section perpendicular to the field~\cite{Tellgren2009}.
This entails computations on excited states of larger molecular systems which become accessible to us with the linear response technique.
Excited states also provide a wider range of possible electronic structures than ground states.

Our implementation is entirely general and is able to handle non-uniform fields which break spin-symmetry, necessitating a two-component representation of orbitals even with a non-relativistic Hamiltonian~\cite{Sen2018a}.
We can thus study how the spin-forbidden excitations behave with a progressive loss in spin-symmetry.
In particular, we study the lowest singlet-triplet transition for various molecules in this paper.  
The behaviour of oscillator strengths is also investigated.

\section{Theory and Implementation}

\subsection{The Hamiltonian}

The non-relativistic Schr\"odinger--Pauli Hamiltonian, which is used in this work, is given by (in atomic units)
\begin{equation}
\hat{H} = \frac{1}{2}\sum_l \hat{\pi_l}^2 - \sum_l v(\mathbf{r}_l) + \sum_{k<l} \frac{1}{r_{kl}} + \sum_l\mathbf{B}_{\text{tot}}(\mathbf{r}_l)\cdot \hat{\mathbf{S}}_l \label{Hamiltonian}
\end{equation}
where $\hat{\boldsymbol{\pi}}_l = -i\nabla_l + \mathbf{A}_{\text{tot}}(\mathbf{r}_l)$ is the mechanical momentum operator.

We choose a linearly varying non-uniform magnetic field, in general, which can be written in the form
\begin{equation}
	\mathbf{B}_{\text{tot}}(\mathbf{r}) = \mathbf{B} + \mathbf{r}_{\mathbf{h}}^T \mathbf{b} - \frac{1}{3}\mathbf{r}_{\mathbf{h}} \, \trace{\mathbf{b}},
\end{equation}
where $\mathbf{B}$ is a uniform (position independent) component, $\mathbf{b}$ is a $3\times 3$ matrix defining the field gradients, and $\mathbf{r}_{\mathbf{h}} = \mathbf{r} - \mathbf{h}$ is the position relative to some reference point $\mathbf{h}$. This form may be viewed as arising from a Taylor expansion around $\mathbf{r} = \mathbf{h}$ truncated at linear order. The corresponding vector potential can be written as
\begin{align}
	\mathbf{A}_{\text{tot}}(\mathbf{r}) &= \frac{1}{2}\mathbf{B} \times \mathbf{r_g}-\frac{1}{3}\mathbf{r_h} \times (\mathbf{r_h}^T\mathbf{b}),
\end{align}
where $\mathbf{r}_{\mathbf{g}} = \mathbf{r} - \mathbf{g}$, $\mathbf{g}$ being the gauge origin. It can be verified that $\mathbf{B}_{\text{tot}} = \nabla\times\mathbf{A}_{\text{tot}}$ and that the magnetic field is divergence free, $\nabla\cdot\mathbf{B}_{\text{tot}} = 0$. In what follows, we quantify the non-uniformity of the field through the anti-symmetric part C$_{\alpha} = \epsilon_{\alpha\beta\gamma} b_{\beta\gamma}$ of the matrix $\mathbf{b}$ and take the symmetric part, $\mathbf{b} = \mathbf{b}^T$, to vanish. We can then write
\begin{align}
	\mathbf{A}_{\text{tot}}(\mathbf{r}) & = \frac{1}{2}\mathbf{B} \times \mathbf{r}_{\mathbf{g}}-\frac{1}{3} \mathbf{r}_{\mathbf{h}} \times (\mathbf{C}\times\mathbf{r}_{\mathbf{h}}),
	\\
	\mathbf{B}_{\text{tot}}(\mathbf{r}) & = \mathbf{B} + \frac{1}{2} \mathbf{C}\times\mathbf{r}_{\mathbf{h}}.
\end{align}
Furthermore, the constant vector encoding the anti-symmetric part of $\mathbf{b}$  equals the curl of the magnetic field, $\nabla \times \mathbf{B}_{\text{tot}} = \mathbf{C}$.

\subsection{Linear Response Formulation}

Due to the loss of time reversal symmetry, Hartree-Fock (HF) computations for atoms and molecules in finite magnetic fields require complex-valued orbitals.
Thus, the exposition below gives a general formulation for complex-valued orbitals without recourse to assumptions of purely real or purely imaginary quantities. A general non-orthonormal basis~\cite{LARSEN_JCP113_8908,CORIANI_JCP126_154108,LUCERO_JCP129_064114,KJAERGAARD_JCP129_054106} (e.g., the atomic orbital basis) is allowed in the derivation and implementation, although the reported applications have been carried out in the orthonormal molecular orbital (MO) basis.

The creation operator $\cop{\sao{1}}$ creates an electron in the spinorbital $\sao{1}$, while the annihilation operator
$\aop{\sao{1}}$ annihilates such an electron. Letting
$S^{\sao{1}\sao{2}}$ denote the inverse of the overlap matrix
$S_{\sao{2}\sao{3}} = \bravac \aop{\sao{2}} \cop{\sao{3}} \vac$, it is
now possible to define
\begin{align}
\copU{\sao{1}} = S^{\sao{1}\sao{2}} \cop{\sao{2}}, \ \aopU{\sao{1}} = \aop{\sao{2}} S^{\sao{2}\sao{1}}.
\end{align}
Note the implicit summation over $\sao{2}$ in the above
expressions. Multiplication by the overlap matrix yields
\begin{align}
\cop{\sao{3}} = S_{\sao{3}\sao{1}} \copU{\sao{1}}, \ \aop{\sao{3}} = \aopU{\sao{1}} S_{\sao{1}\sao{3}}.
\end{align}

Borrowing terminology from differential geometry, indices occur in
both \emph{covariant} (subscript) and \emph{contravariant}
(superscript) positions~\cite{HELGAKER_AQC19_183,HEADGORDON_JCP108_616}. We rely on the summation convention that
indices that occur in both positions are summed over, unless otherwise
indicated. Unitary or orbital invariance is ensured when all
contractions are of this form. In general, contraction with the
overlap matrix or its inverse lowers and raises indices, respectively,
in the manner seen above. Clearly, the distinction between covariant
and contravariant indices disappears in an orthonormal basis, where both the
overlap matrix and its inverse equals the identity matrix.

We also note that a generic second-quantized 1-particle
operator $\op{A}$ has the form
\begin{align}
\op{A} = A^{\sao{1}\sao{2}} \cop{\sao{1}} \aop{\sao{2}}.
\end{align}
By contrast, a generic 1-particle reduced density operator is of the form
\begin{align}
\op{D} = \cop{\sao{1}} \vac D^{\sao{1}\sao{2}} \bravac \aop{\sao{2}}.
\end{align}
A linear map
\begin{equation}
 \begin{split}
   \mathcal{L}_{1\to \mathrm{F}}(\op{R}) & = R^{\sao{2}\sao{1}} \cop{\sao{2}} \aop{\sao{1}}, \quad \text{with}
     \\
  R^{\sao{2}\sao{1}} & = S^{\sao{2}\sao{3}} \bravac
  \aop{\sao{3}} \op{A} \cop{\sao{4}} \vac S^{\sao{4}\sao{1}},
 \end{split}
\end{equation}
converts an operator of the second form into the first form. That is, $R^{\sao{2}\sao{1}} \cop{\sao{2}} \vac \bravac \aop{\sao{1}}$ is mapped to $R^{\sao{2}\sao{1}} \cop{\sao{2}} \aop{\sao{1}}$.

Now let $\ket{\text{gs}}$ be the exact ground state and $\ket{X} = \op{X}
\ket{\text{gs}}$ be an exact excited state, generated using the
corresponding excitation operator $\op{X}$. Then
\begin{align}
[\op{H}, \op{X}] \ket{\text{gs}} = \omega \op{X} \ket{\text{gs}},
\end{align}
with
\begin{align}
\omega = \bra{X} \op{H} \ket{X} - \bra{\text{gs}} \op{H} \ket{\text{gs}}
= E_X - E_{\text{gs}}
\end{align}
being the excitation energy. Since $(\op{H} - E_X) \ket{X} = 0$, it also
follows that
\begin{align}\label{eqRPASTART}
\bra{\text{gs}} \op{E}_{\sao{5}\sao{6}} (\op{H} - E_X) \ket{X} = 0
\end{align}
for any operator $\op{E}_{\sao{5}\sao{6}} = \cop{\sao{5}} \aop{\sao{6}}$. Adding and subtracting
$\bra{\text{gs}} \op{H} \op{E}_{\sao{5}\sao{6}} \ket{X} = E_{\text{gs}} \bra{\text{gs}} \op{E}_{\sao{5}\sao{6}} \ket{X}$ to
Eq.~\eqref{eqRPASTART} yields
\begin{align}
\bra{\text{gs}} [\op{E}_{\sao{5}\sao{6}}, \op{H}] \ket{X} - \omega \bra{\text{gs}} \op{E}_{\sao{5}\sao{6}} \ket{X} & = 0.
\end{align}
Exploiting the fact that $\bra{\text{gs}} \op{X} = 0$, the
equation may be rewritten further to take the form
\begin{align}
\bra{\text{gs}} [[\op{H}, \op{E}_{\sao{5}\sao{6}}], \op{X}] \ket{\text{gs}} = \omega \bra{\text{gs}} [\op{X}, \op{E}_{\sao{5}\sao{6}}] \ket{\text{gs}}
\end{align}

At this point three assumptions are made. Firstly, the expectation value
with respect to the ground state $\ket{\text{gs}}$ is replaced by the
expectation value with respect to the HF state
$\ket{\text{HF}}$. Secondly, the excitation operator $\op{X}$ is
assumed to involve only single excitations (and deexcitations),
\begin{align}
  \label{eqRPAXform}
\op{X} = X^{\sao{1}\sao{2}} \op{E}_{\sao{1}\sao{2}}.
\end{align}
Thirdly, we decompose $\op{X} = \op{Y} + \op{Z}$ into an excitation ($\op{Y}$) and a deexcitation ($\op{Z}$) component with respect to
$\ket{\text{HF}}$.
In order to remove redundant component of $\op{X}$ we apply a
projection to the occupied orbitals in the 1-electron sector of Fock
space. To get back to an operator of the form in
Eq.~\eqref{eqRPAXform} that acts on the whole Fock space, we require the linear map $\mathcal{L}_{1\to \mathrm{F}}$. The non-redundancy conditions may thus be written
\begin{equation}
  \op{X} =  \mathcal{L}_{1\to \mathrm{F}}(\op{Q} \op{X} \op{P} + \op{P} \op{X} \op{Q})
    = \mathcal{L}_{1\to \mathrm{F}}(\op{Q} \op{Y} \op{P} + \op{P} \op{Z} \op{Q}),
\end{equation}
or
\begin{align}
\op{Y} = \mathcal{L}_{1\to \mathrm{F}}( \op{Q} \op{Y} \op{P} ),
\quad
\op{Z} = \mathcal{L}_{1\to \mathrm{F}}( \op{P} \op{Z} \op{Q} ),
\end{align}
where $\op{P} = \cop{\sao{2}} \vac
P^{\sao{2}\sao{1}} \bravac \aop{\sao{1}}$ denotes the (1-particle
reduced) density operator for the Hartree--Fock state and $\op{Q} =
\op{I} - \op{P}$, with $\op{I} = \cop{\sao{2}} \vac S^{\sao{2}\sao{1}}
\bravac \aop{\sao{1}}$ the identity operator within the
one-electron sector.
Alternatively, we could have used an $N$-electron projector
$\mathcal{P} = \ket{\text{HF}} \bra{\text{HF}}$ and defined an
analogous map $\mathcal{L}_{N\to \mathrm{F}}$ from the $N$-electron sector to the full
Fock space.
Projection of redundant degrees of freedom is crucial
for avoiding spurious solutions~\cite{LUCERO_JCP129_064114}.

From these three assumptions, it now follows that,
\begin{align} 
\label{eqRPAHF}
\bra{\text{HF}} [[\op{H}, \op{E}_{\sao{5}\sao{6}}], \op{X}] \ket{\text{HF}} = \omega \bra{\text{HF}} [\op{X}, \op{E}_{\sao{5}\sao{6}}] \ket{\text{HF}}.
\end{align}
Our working equations are obtained by exploiting commutation relations such as
\begin{align}
	\label{eqCOPEcom}
	[\cop{\sao{1}}, \op{E}_{\sao{2}\sao{3}}] & = \cEcom{\sao{1}}{\sao{2}}{\sao{3}}, \\
	\label{eqAOPEcom}
	[\aop{\sao{1}}, \op{E}_{\sao{2}\sao{3}}] & = \aEcom{\sao{1}}{\sao{2}}{\sao{3}}, \\
	\label{eqEEcom}
	[\op{E}_{\sao{1}\sao{2}}, \op{E}_{\sao{3}\sao{4}}] & = \EEcom{\sao{1}}{\sao{2}}{\sao{3}}{\sao{4}},
\end{align}
where indices that refer to creation (annihilation) operators on the left-hand side must do so also on the right-hand side.  This holds also for the indices in the overlap matrix, since $S_{\sao{1}\sao{2}} = \bravac \aop{\sao{1}} \cop{\sao{2}} \vac$.
Writing $\op{e}_{\sao{1}\sao{2}\sao{3}\sao{4}} = \cop{\sao{1}}
\cop{\sao{3}} \aop{\sao{4}} \aop{\sao{2}}$ for a two-electron operator
string in the Hamiltonian, it is also notable that
\begin{align}
	[\op{e}_{\sao{1}\sao{2}\sao{3}\sao{4}}, \op{E}_{\sao{5}\sao{6}}] & =
	\eEcom{\sao{1}}{\sao{2}}{\sao{3}}{\sao{4}}{\sao{5}}{\sao{6}}.
\end{align}

Using Eq.~\eqref{eqEEcom}, the right-hand side of Eq.~\eqref{eqRPAHF} may be written as
\begin{align}
\label{eqMETRICTRANSF}
& \omega S^{[2]}_{\sao{5}\sao{6},\sao{7}\sao{8}} X^{\sao{7}\sao{8}} \nonumber\\
& = \omega \bra{\text{HF}} [\op{X}, \op{E}_{\sao{5}\sao{6}}] \ket{\text{HF}}
= \omega \bra{\text{HF}} \EEcom{\sao{7}}{\sao{8}}{\sao{5}}{\sao{6}} \ket{\text{HF}} X^{\sao{7}\sao{8}}
\nonumber\\
& = \omega (P_{\sao{6}\sao{7}} X^{\sao{7}\sao{8}} S_{\sao{8}\sao{5}} - S_{\sao{6}\sao{7}} X^{\sao{7}\sao{8}} P_{\sao{8}\sao{5}}) = \omega [P,X]_{\sao{6}\sao{5}}.
\end{align}

In a non-orthonormal basis the Hamiltonian takes the form
\begin{align}
\op{H} & = h^{\sao{1}\sao{2}}  \op{E}_{\sao{1}\sao{2}} + \frac{1}{2} g^{\sao{1}\sao{2}\sao{3}\sao{4}} e_{\sao{1}\sao{2}\sao{3}\sao{4}} \nonumber 
\end{align}
where $g^{\sao{1}\sao{2}\sao{3}\sao{4}} = S^{\sao{1}\sao{5}}
S^{\sao{6}\sao{2}} g_{\sao{5}\sao{6}\sao{7}\sao{8}} S^{\sao{3}\sao{7}}
S^{\sao{8}\sao{4}}$. Decomposing the Hamiltonian into its one- and two-electron parts, $\op{H} = \op{h} +
\op{g}$, it now follows from the
above commutation relations that,
\begin{align}
& [\op{h},\op{E}_{\sao{5}\sao{6}}] = h^{\sao{1}\sao{2}} ( \EEcom{\sao{1}}{\sao{2}}{\sao{5}}{\sao{6}} ) = \tUD{h}{\sao{1}}{\sao{5}} \op{E}_{\sao{1}\sao{6}} - \tDU{h}{\sao{6}}{\sao{2}} \op{E}_{\sao{5}\sao{2}} \\
& [\op{g},\op{E}_{\sao{5}\sao{6}}] \nonumber \\
& = \frac{1}{2} ( \tUUUD{g}{\sao{1}}{\sao{2}}{\sao{3}}{\sao{5}}  \op{e}_{\sao{1}\sao{2}\sao{3}\sao{6}} + \tUDUU{g}{\sao{1}}{\sao{5}}{\sao{3}}{\sao{4}} \op{e}_{\sao{1}\sao{6}\sao{3}\sao{4}} - \tDUUU{g}{\sao{6}}{\sao{2}}{\sao{3}}{\sao{4}} \op{e}_{\sao{5}\sao{2}\sao{3}\sao{4}} - \tUUDU{g}{\sao{1}}{\sao{2}}{\sao{6}}{\sao{4}} \op{e}_{\sao{1}\sao{2}\sao{5}\sao{4}}).
\end{align}

In response theory the double commutator $[[\op{H},
\op{E}_{\sao{5}\sao{6}}], \op{E}_{\sao{7}\sao{8}}]$ appears as a
central quantity.
After a tedious but straightforward calculation, the final simplified
form is found to be
\begin{align}
& [[\op{g},\op{E}_{\sao{5}\sao{6}}],\op{E}_{\sao{7}\sao{8}}] \nonumber \\
& = \tUUUD{g}{\sao{1}}{\sao{2}}{\sao{3}}{\sao{5}} S_{\sao{6}\sao{7}} \op{e}_{\sao{1}\sao{2}\sao{3}\sao{8}} + \tUDUD{g}{\sao{1}}{\sao{7}}{\sao{3}}{\sao{5}} \op{e}_{\sao{1}\sao{8}\sao{3}\sao{6}} - \tDUUD{g}{\sao{8}}{\sao{2}}{\sao{3}}{\sao{5}} \op{e}_{\sao{7}\sao{2}\sao{3}\sao{6}} - \tUUDD{g}{\sao{1}}{\sao{2}}{\sao{8}}{\sao{5}} \op{e}_{\sao{1}\sao{2}\sao{7}\sao{6}}
\nonumber\\
& \ - \tDUUD{g}{\sao{6}}{\sao{2}}{\sao{3}}{\sao{7}} \op{e}_{\sao{5}\sao{2}\sao{3}\sao{8}} - \tDDUU{g}{\sao{6}}{\sao{7}}{\sao{3}}{\sao{4}} \op{e}_{\sao{5}\sao{8}\sao{3}\sao{4}} + \tDUUU{g}{\sao{6}}{\sao{2}}{\sao{3}}{\sao{4}} S_{\sao{8}\sao{5}} \op{e}_{\sao{7}\sao{2}\sao{3}\sao{4}} + \tDUDU{g}{\sao{6}}{\sao{2}}{\sao{8}}{\sao{4}} \op{e}_{\sao{5}\sao{2}\sao{7}\sao{4}}
\end{align}
Within Hartree--Fock response theory, the tensor
$W^{[2]}_{\sao{5}\sao{6},\sao{7}\sao{8}} = \bra{\text{HF}}
[[\op{H},\op{E}_{\sao{5}\sao{6}}],\op{E}_{\sao{7}\sao{8}}]
\ket{\text{HF}}$ plays the role of a Hessian for the electronic
degrees of freedom. For the one-electron part one immediately obtains
\begin{align}
& W^{[2,1\mathrm{el}]}_{\sao{5}\sao{6},\sao{7}\sao{8}} = \bra{\text{HF}} [[\op{h},\op{E}_{\sao{5}\sao{6}}],\op{E}_{\sao{7}\sao{8}}] \ket{\text{HF}}
\nonumber\\
& = \tUD{h}{\sao{1}}{\sao{5}} (S_{\sao{6}\sao{7}} P_{\sao{8}\sao{1}} - S_{\sao{8}\sao{1}} P_{\sao{6}\sao{7}}) - \tDU{h}{\sao{6}}{\sao{2}} (S_{\sao{2}\sao{7}} P_{\sao{8}\sao{5}} - S_{\sao{8}\sao{5}} P_{\sao{2}\sao{7}}),
\end{align}
where it is being exploited that $P_{\sao{2}\sao{1}} = \bra{\text{HF}}
\op{E}_{\sao{1}\sao{2}} \ket{\text{HF}}$.
To compute the two-electron
part it is useful to first note that the 2-particle reduced density
matrix $\Gamma_{\sao{2}\sao{1}\sao{4}\sao{3}} = \bra{\text{HF}}
\cop{\sao{1}} \cop{\sao{3}} \aop{\sao{4}} \aop{\sao{2}}
\ket{\text{HF}}$ of a Slater-determinantal state satisfies
\begin{align}
\Gamma_{\sao{2}\sao{1}\sao{4}\sao{3}} = \GammaExp{\sao{2}}{\sao{1}}{\sao{4}}{\sao{3}}.
\end{align}
and also identify $\tUUUD{g}{\sao{1}}{\sao{2}}{\sao{3}}{\sao{5}} P_{\sao{2}\sao{1}}$ as the Coulomb matrix element
$\tUD{J}{\sao{3}}{\sao{5}} = \tUD{J}{\sao{3}}{\sao{5}}(P)$ and
similarly for other combinations of covariant and contravariant
indices as well as for exchange contractions. In general, contraction
of a single density matrix with the first (or last) two indices of the
g-tensor yields a Coulomb matrix, while contraction of the middle (or
first and last) indices yield an exchange matrix. We thus obtain,
\begin{equation}
 \begin{split}
 W^{[2,2\mathrm{el}]}_{\sao{5}\sao{6},\sao{7}\sao{8}} 
& = S_{\sao{6}\sao{7}} P_{\sao{8}\sao{1}} \tUD{G}{\sao{1}}{\sao{5}}
+ \tDU{G}{\sao{6}}{\sao{2}} P_{\sao{2}\sao{7}} S_{\sao{8}\sao{5}} - P_{\sao{6}\sao{7}} G_{\sao{8}\sao{5}} - G_{\sao{6}\sao{7}} P_{\sao{8}\sao{5}}
    \\
& \ \ - P_{\sao{6}\sao{3}} P_{\sao{2}\sao{7}} \tDUUD{g}{\sao{8}}{\sao{2}}{\sao{3}}{\sao{5}}
+ P_{\sao{6}\sao{1}} P_{\sao{2}\sao{7}} \tUUDD{g}{\sao{1}}{\sao{2}}{\sao{8}}{\sao{5}}
     \\
& \ \ - P_{\sao{2}\sao{5}} \tDUUD{g}{\sao{6}}{\sao{2}}{\sao{3}}{\sao{7}} P_{\sao{8}\sao{3}}
+ P_{\sao{4}\sao{5}} \tDDUU{g}{\sao{6}}{\sao{7}}{\sao{3}}{\sao{4}} P_{\sao{8}\sao{3}}
    \\
& \ \  + \tDUDU{g}{\sao{6}}{\sao{2}}{\sao{8}}{\sao{4}}
(\GammaExp{\sao{2}}{\sao{5}}{\sao{4}}{\sao{7}})
    \\
 & + \tUDUD{g}{\sao{1}}{\sao{7}}{\sao{3}}{\sao{5}}
(\GammaExp{\sao{8}}{\sao{1}}{\sao{6}}{\sao{3}})
 \end{split}
\end{equation}
where, $G_{\sao{2}\sao{1}}(P) = J_{\sao{2}\sao{1}}(P) - K_{\sao{2}\sao{1}}(P)$.

Next, it is useful to derive the transformation of an orbital
rotation operator $\op{X} = X^{\sao{7}\sao{8}} \op{E}_{\sao{7}\sao{8}}$ by the Hessian. The resulting index contractions actually simplify the result considerably,
\begin{align}
\label{eqHESSIANTRANSF2}
 W^{[2]}_{\sao{5}\sao{6},\sao{7}\sao{8}} X^{\sao{7}\sao{8}} & = W^{[2,1\mathrm{el}]}_{\sao{5}\sao{6},\sao{7}\sao{8}} X^{\sao{7}\sao{8}} + W^{[2,2\mathrm{el}]}_{\sao{5}\sao{6},\sao{7}\sao{8}} X^{\sao{7}\sao{8}} \nonumber \\
& = -[[P,X], F(P)]_{\sao{6}\sao{5}} - [P, G([P,X])]_{\sao{6}\sao{5}},
\end{align}
where $F_{\sao{2}\sao{1}} = F_{\sao{2}\sao{1}}(P) = h_{\sao{2}\sao{1}}
+ G_{\sao{2}\sao{1}}(P)$ is the Fock matrix computed from the density
matrix $P^{\sao{1}\sao{2}}$ and $G_{\sao{2}\sao{1}}([P,X]) =
J_{\sao{2}\sao{1}}([P,X]) - K_{\sao{2}\sao{1}}([P,X])$ is the Coulomb
and exchange contributions computed from the density matrix
$[P,X]^{\sao{2}\sao{1}}$.

Finally, it is useful to express the metric and Hessian
transformations in the conventional form (covariant overlap and Fock
matrices, contravariant density and orbital rotation matrices),
\begin{align}
\label{eqMETRICIMPL}
\omega S^{[2]}_{\sao{5}\sao{6},\sao{7}\sao{8}} X^{\sao{7}\sao{8}} & = \omega (S_{\sao{6}\sao{1}} P^{\sao{1}\sao{2}} S_{\sao{2}\sao{7}} X^{\sao{7}\sao{8}} S_{\sao{8}\sao{5}} - S_{\sao{6}\sao{7}} X^{\sao{7}\sao{8}} S_{\sao{8}\sao{1}} P^{\sao{1}\sao{2}} S_{\sao{2}\sao{5}}) \nonumber \\
& = \omega S_{\sao{6}\sao{1}} [P,X]^{\sao{1}\sao{2}} S_{\sao{2}\sao{5}}, \\
 W^{[2]}_{\sao{5}\sao{6},\sao{7}\sao{8}} X^{\sao{7}\sao{8}} & = -S_{\sao{6}\sao{1}} [[P,X], F(P)]^{\sao{1}\sao{2}} S_{\sao{2}\sao{5}} \nonumber\\
   & \ \ - S_{\sao{6}\sao{1}} [P, G([P,X])]^{\sao{1}\sao{2}} S_{\sao{2}\sao{5}}
\end{align}
where not all commutators have been written out in full.

The structure of the excitation operator depends on the Hartree--Fock state used as reference. 
In the restricted Hartree--Fock model ($\ket{\text{HF}} = \ket{\text{RHF}}$), $\op{X}$ is a spin-free/spin-summed operator labelled by spatial orbitals only.
We can thus generate only singlet excited states with an RHF reference.
It is possible to generate triplet states by choosing $\op{X}$ to be a triplet coupled combination of the spin-free operators but this is not implemented in our case.
When the reference is an unrestricted Hartree--Fock state ($\ket{\text{HF}} = \ket{\text{UHF}}$), $\op{X}$ is a set of spin-conserving excitations labelled by spinorbitals.
However, a triplet state with $m_S=0$ may be generated from a singlet UHF reference as the combining co-efficients, $c$, may converge to give $c_{\uparrow\uparrow} =c_{\downarrow\downarrow}$. With UHF references of other spin multiplicities and $m_S$ values, we can generate various other spin multiplicities.
The most general reference state is the General Hartree--Fock state ($\ket{\text{HF}} = \ket{\text{GHF}}$) made of two-component orbitals.
This allows the flexibility of generating a spin-mixed state in the presence of a non-uniform magnetic field when $S^2$ ceases to be a good quantum number.
For further details on the GHF method we refer to our earlier publication~\cite{Sen2018a}.
In this case, $\hat{X}$ is also a two-component excitation operator having $\uparrow\uparrow$, $\uparrow\downarrow$, $\downarrow\uparrow$ and $\downarrow\downarrow$ components.

Transition moments can be simply derived by considering general transition matrix elements of one electron operators.
Let $\op{A} = A^{\sao{1}\sao{2}} \cop{\sao{1}} \aop{\sao{2}}$ be an
arbitrary 1-particle operator and $\op{X}_k$ the $k^{\text{th}}$ RPA excitation operator. Then
\begin{equation}
 \begin{split}
\bra{\text{HF}} \op{A} \ket{X_k} & = \bra{\text{HF}} \op{A} \op{X}_k \ket{\text{HF}} 
  = \bra{\text{HF}} [\op{A}, \op{X}_k] \ket{\text{HF}},
 \end{split}
\end{equation}
or, in terms of the density matrix,
\begin{equation}
 \begin{split}
\bra{\text{HF}} \op{A} \ket{X_k} & = \trace{\op{P} [\op{A}, \op{X}_k]} = \trace{[\op{A}, \op{P}] \op{X}_k}
       \\
    &  = \trace{\op{A} [\op{P}, \op{X}_k]}.
 \end{split}
\end{equation}
Thus, the transition density operator for the $k^{\text{th}}$ state can be identified as
\begin{equation}
 \begin{split}
	\op{\mathcal{M}}_{k0} & =  [\op{P}, \op{X}_k] = \op{P} \op{X}_k \op{Q} - \op{Q} \op{X}_k \op{P}.
 \end{split}
\end{equation}
and a transition property can then be evaluated by the expression
\begin{equation}
	\bra{\text{HF}} \op{A} \ket{X_k} = \trace{\op{A} \op{\mathcal{M}}_{k0}}.
\end{equation}

There are several equivalent formulas for computing the oscillator strength $f$ for an electric dipole transition~\cite{Chandrasekhar1945}.
The two most commonly used are the dipole length formula,
\begin{equation}
	f_l = 2\Delta E \Big| \bra{0} \sum_{i=1}^{N} \mathbf{r}_{i}\ket{X_k} \Big|^2,
\end{equation}
and the dipole velocity formula,
\begin{equation}
   f_v = \frac{2}{\Delta E} \Big| \bra{0} \sum_{i=1}^{N} \op{\boldsymbol{\pi}}_i \ket{X_k} \Big|^2
\end{equation}
for a transition of energy $\Delta E$ from $\ket{0}$ to $\ket{X_k}$
in an $N$-electron system. Note that mechanical momentum operator
$\op{\boldsymbol{\pi}} = \op{\mathbf{p}} + \mathbf{A}_{\mathrm{tot}}$,
as opposed to the canonical momentum operator $\op{\mathbf{p}} =
-i\nabla$, appears in the velocity gauge at non-zero fields $\mathbf{A}_{\mathrm{tot}}$.
For exact wave functions the two values must agree but in approximate
descriptions they will typically be different~\cite{Harris1969}.
In the complete basis set limit, RPA is one of the few approximate theories which maintains the equivalence~\cite{Harris1969,Dalgaard1980,Jorgensen1983}.
While an explicit proof for the equivalence is available for real orbitals~\cite{Jorgensen1975}, a proof for complex orbitals (as is necessarily used in our studies) is not available in the literature. In our results and discussions section we have numerically demonstrated that the equivalence holds for complex orbitals as well.

\subsection{Implementation}

In the linear response equations,
\begin{align}
(W^{[2]}_{\sao{5}\sao{6},\sao{7}\sao{8}} - \omega S^{[2]}_{\sao{5}\sao{6},\sao{7}\sao{8}}) X^{\sao{7}\sao{8}} = -B^{[1]}_{\sao{5}\sao{6}}.
\end{align}
the frequency $\omega$ is an
arbitrary externally given parameter, whereas it is an initially
unknown excitation energy in the RPA equation. Apart from this
difference, the RPA equation corresponds to the special case of a
vanishing property gradient $B^{[1]}_{\sao{5}\sao{6}} = 0$.
This equation is solved by a modified Davidson method involving Krylov iterations with some additional considerations to account for the metric $S^{[2]}$. 

Suppose that $m$ trial vectors
$\op{b}^{(i)}$, $1 \leq i \leq m$, have already been analyzed. For
each of these trial vectors there is a Hermitian-conjugate partner
$\op{b}^{(i)\dagger}$, the metric and Hessian transformation of which
is closely related to those of $\op{b}^{(i)}$. As an {\it Ansatz} for
a new approximate orbital rotation operator $\op{X}$, the following
linear expansion is chosen
\begin{align}
\op{X} & = \sum_{i=1}^m d_i \op{b}^{(i)} + \sum_{i=1}^m d_{m+i} \op{b}^{(i)\dagger} = \sum_{i=1}^{2m} d_i \op{b}^{(i)},
\end{align}
where the convention $\op{b}^{(m+i)} = \op{b}^{(i)\dagger}$, $1 \leq i
\leq m$, has been introduced to get to the last expression. The
problem of determining a new trial vector has now been reduced to
determining $2m$ coefficients $d_i$. The metric and Hessian
transformation are linear so that
\begin{align}
S^{[2]}_{\sao{5}\sao{6},\sao{7}\sao{8}} X^{\sao{7}\sao{8}} & = \sum_{i=1}^{2m} d_i (S^{[2]}_{\sao{5}\sao{6},\sao{7}\sao{8}} b^{(i);\sao{7}\sao{8}}) = \sum_{i=1}^{2m} d_i s^{(i)}_{\sao{5}\sao{6}}, \\
W^{[2]}_{\sao{5}\sao{6},\sao{7}\sao{8}} X^{\sao{7}\sao{8}} & = \sum_{i=1}^{2m} d_i (W^{[2]}_{\sao{5}\sao{6},\sao{7}\sao{8}} b^{(i);\sao{7}\sao{8}}) = \sum_{i=1}^{2m} d_i w^{(i)}_{\sao{5}\sao{6}}
\end{align}
Projection of $(W^{[2]}_{\sao{5}\sao{6},\sao{7}\sao{8}} - \omega
S^{[2]}_{\sao{5}\sao{6},\sao{7}\sao{8}}) X^{\sao{7}\sao{8}} = 0$ onto
an arbitrary trial vector $\op{b}^{(l)}$, $1 \leq l \leq 2m$, now
gives
\begin{align}
&b^{(l);\sao{5}\sao{6}} (W^{[2]}_{\sao{5}\sao{6},\sao{7}\sao{8}} - \omega
S^{[2]}_{\sao{5}\sao{6},\sao{7}\sao{8}}) X^{\sao{7}\sao{8}} \nonumber \\
& = \sum_{i=1}^{2m} (W^{\text{red}}_{li} - \omega S^{\text{red}}_{li}) d_i = 0,
\quad \text{for all}\ 1 \leq l \leq 2m,
\end{align}
with the obvious definitions of the reduced-space matrices
$S^{\text{red}}_{li}$ and $W^{\text{red}}_{li}$. Solution of the
reduced-space generalized eigenvalue equation yields the coefficients
$d_i$ and the eigenvalue $\omega$. Once these are determined, the residual
\begin{align}
R_{\sao{6}\sao{5}} & = W^{[2]}_{\sao{5}\sao{6},\sao{7}\sao{8}} - \omega S^{[2]}_{\sao{5}\sao{6},\sao{7}\sao{8}}
\end{align}
can be computed as a measure of how large errors remain in the
approximate $\op{X}$. If the error is too large, the residual and its
Hermitian-conjugate may be added as new trial vectors.

Let $m$ trial vectors $\op{b}^{(i)}$ and their Hermitian-conjugates
$\op{b}^{(m+i)} = \op{b}^{(i)\dagger}$ be given. We now wish to define
new linear combinations, $1 \leq l \leq 2m$,
\begin{align}
\op{c}^{(l)} & = \sum_{i=1}^{2m} x_{il} \op{b}^{(i)} = \sum_{i=1}^{m} x_{il} \op{b}^{(i)} + \sum_{i=1}^{m} x_{m+i,l} \op{b}^{(i)\dagger}.
\end{align}
The new trial vectors should be orthonormal with respect to the
natural scalar product between operators,
\begin{align}
(\op{A}, \op{B}) & = \mathrm{Tr}(\op{A}^\dagger \op{B}).
\end{align}
Furthermore, the new trial vectors
should come in Hermitian-conjugated pairs. Thus,
\begin{align}
\mathrm{Tr}(\op{c}^{(k)\dagger} \op{c}^{(l)}) & = \delta_{kl},  \quad 1 \leq k,l \leq 2m, \\
\op{c}^{(l)\dagger} & = \op{c}^{(m+l)}, \quad 1 \leq l \leq m.
\end{align}

Subsequently, the Hessian and metric transformation of the new trial vectors must be computed.
This requires the computation of $G([P,X])$, Coulomb and exchange integrals with a non-Hermitian density matrix.
Even when the MO basis is used, the computation of the two-electron integrals continues to be in the AO basis in our implementation leading to additional computational cost for MO to AO and back transformations for the density matrix and integrals respectively. This constitutes the rate determining step of the response computation.

Note that the above expression yields the covariant matrix
elements of the residual, whereas the trial vectors have been
expressed in terms of contravariant matrix elements. The matrix
elements that represent the new trial vector are therefore
\begin{align}
R^{\sao{3}\sao{4}} = S^{\sao{3}\sao{6}} R_{\sao{6}\sao{5}} S^{\sao{5}\sao{4}}.
\end{align}
In addition, the residual may be orthogonalized against the previous
trial vectors.
Note that the covariant-to-contravariant transformation is not always clearly distinguished from preconditioning and sometimes omitted entirely. For example, Coriani {\it et al.}~\cite{CORIANI_JCP126_154108} had no analogue of this step. A follow up work by Kj\ae rgaard {\it et al}.~\cite{KJAERGAARD_JCP129_054106} explicitly included the transformation, while discussing it using a different terminology.
In our view, the covariant-to-contravariant transformation seems to be necessary while,  preconditioning is something that can be done in addition. 

Finally, it is important to enforce the condition
\begin{align}
\label{eqRPAEXCSTRUCTURE}
\op{X} = \mathcal{L}_{1\to \mathrm{F}}(\op{P} \op{X} \op{Q} + \op{Q} \op{X} \op{P})
\end{align}
during the iterative solution process. This may be done by simply
replacing any new trial vector $\op{b}^{(k)}$ by $\mathcal{L}_{1\to \mathrm{F}}(\op{P}
\op{b}^{(k)} \op{Q} + \op{Q} \op{b}^{(k)} \op{P})$.
We have implemented a simple preconditioner for the MO basis, transforming the residual as
\begin{align}
R'_{ia} = \frac{R_{ia}}{\epsilon_a - \epsilon_i - \omega}, \ R'_{ai} = \frac{R_{ai}}{\epsilon_a - \epsilon_i + \omega},
\end{align}
where $i$ denotes an occupied MO and $a$ denotes an unoccupied MO.

\section{Results and Discussions}

The most dramatic effect of strong magnetic fields is the change in the ground state of molecules.
With increase in the strength of the field, states of higher spin multiplicities rapidly come down and become the ground state, even for closed shell molecules.
In our response calculations, we have ensured that we follow the same reference state throughout the full range of magnetic fields studied, allowing negative excitation energies, if necessary.
Equilibrium geometries of molecules also change in the presence of a magnetic field. In our computations we have fixed the geometries at the zero-field values except where specifically mentioned.
When an excited state falls below the zero-field ground state, one must ideally optimize the geometry of the excited state at every field strength also in order to predict the exact crossover point, since both response calculations and EOMCC calculations give adiabatic excited states from a reference ground state. However, since we are aiming at a qualitative understanding, it was not necessary to obtain exact crossover points for the purpose of this paper.
As we go to larger molecules the same effects are seen at weaker fields.
The basis sets employed in this study come from the family of Dunning's correlation consistent basis sets~\cite{DUNNING_JCP90_1989,WOON_JCP100_2975} with augmentation with diffuse functions. 
The names of the basis sets are prefixed with `L' to denote the use of London atomic orbitals and `u' to indicate that the basis sets are uncontracted. 

\subsection{Behaviour of Excited States with Changing Magnetic Fields}

States may be classified as diamagnetic or paramagnetic based on
the energetic response to an applied field. For (non-degenerate) ground states the
leading order term is usually second order in the field, so that
diamagnetism (paramagnetism) becomes associated with negative
(positive) values on the diagonal of the magnetizability tensor. By contrast, excited
states are often degenerate at zero field and can consequently have a permanent orbital or spin magnetic dipole moment,
leading to a linear effect on the energy. Hence, for excited states we often need to distinguish between first order and second order dia-/paramagnetism.

In what follows, the spin quantization axis is taken to be parallel/anti-parallel to a uniform $\mathbf{B}$.
Since the spin-Zeeman term is linear in $\mathbf{B}$, it dominates the weak-field response for non-singlet states with $m_S \neq 0$.
The spin-Zeeman energy, $\frac{1}{2} |\mathbf{B}| m_S$, is included in the reported energies of excited states with $m_S \neq 0$.
The orbital effects, on the other hand, have both linear and quadratic components and will dominate at sufficiently strong fields.
An interesting situation arises for molecules with second order paramagnetism in fields perpendicular to the bond axis (linear molecules) or plane (planar molecules with $\Pi$ electrons), where the orbital-Zeeman interaction turns out to be paramagnetic even for $\Sigma$ states with $m_L= 0$.
This behaviour may be traced to strong coupling with excited states at weak fields.
A transition to normal diamagnetic behaviour occurs at stronger fields.
Closed shell paramagnetic molecules are characterized by a small HOMO-LUMO gap.
Some well-known small examples which have been studied in this paper are BH, CH$^+$ and C$_4$H$_4$.

\begin{figure}
	\centering
	\includegraphics[width=\linewidth]{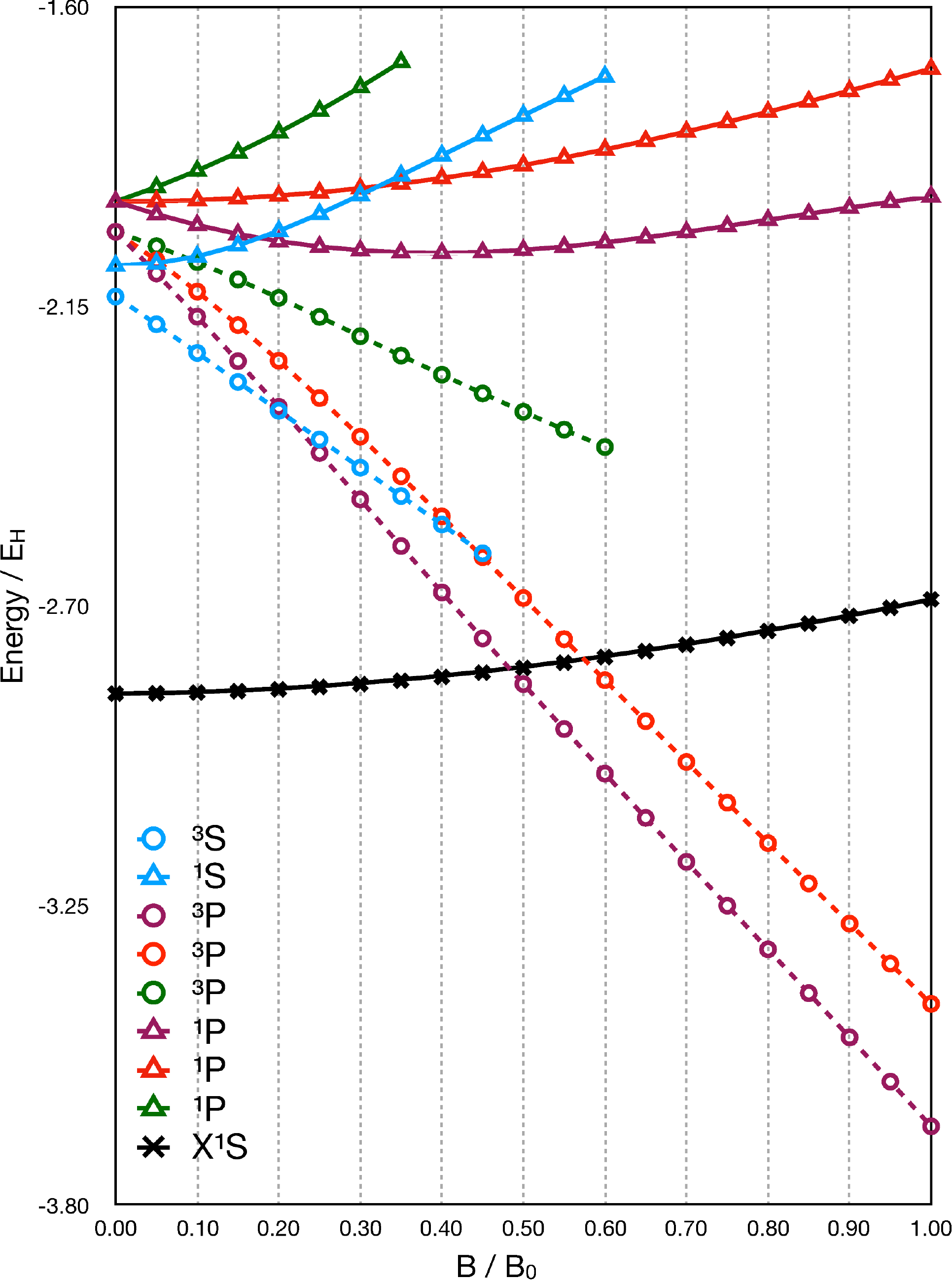}
	\caption{Spectrum of the He atom subject to uniform magnetic fields. The ground and excited states are computed using Hartree-Fock (GHF) and RPA respectively with the Luaug-cc-pCVQZ basis.}
	\label{fig:heexcbfig1}
\end{figure}

\begin{figure}
	\centering
	\includegraphics[width=\linewidth]{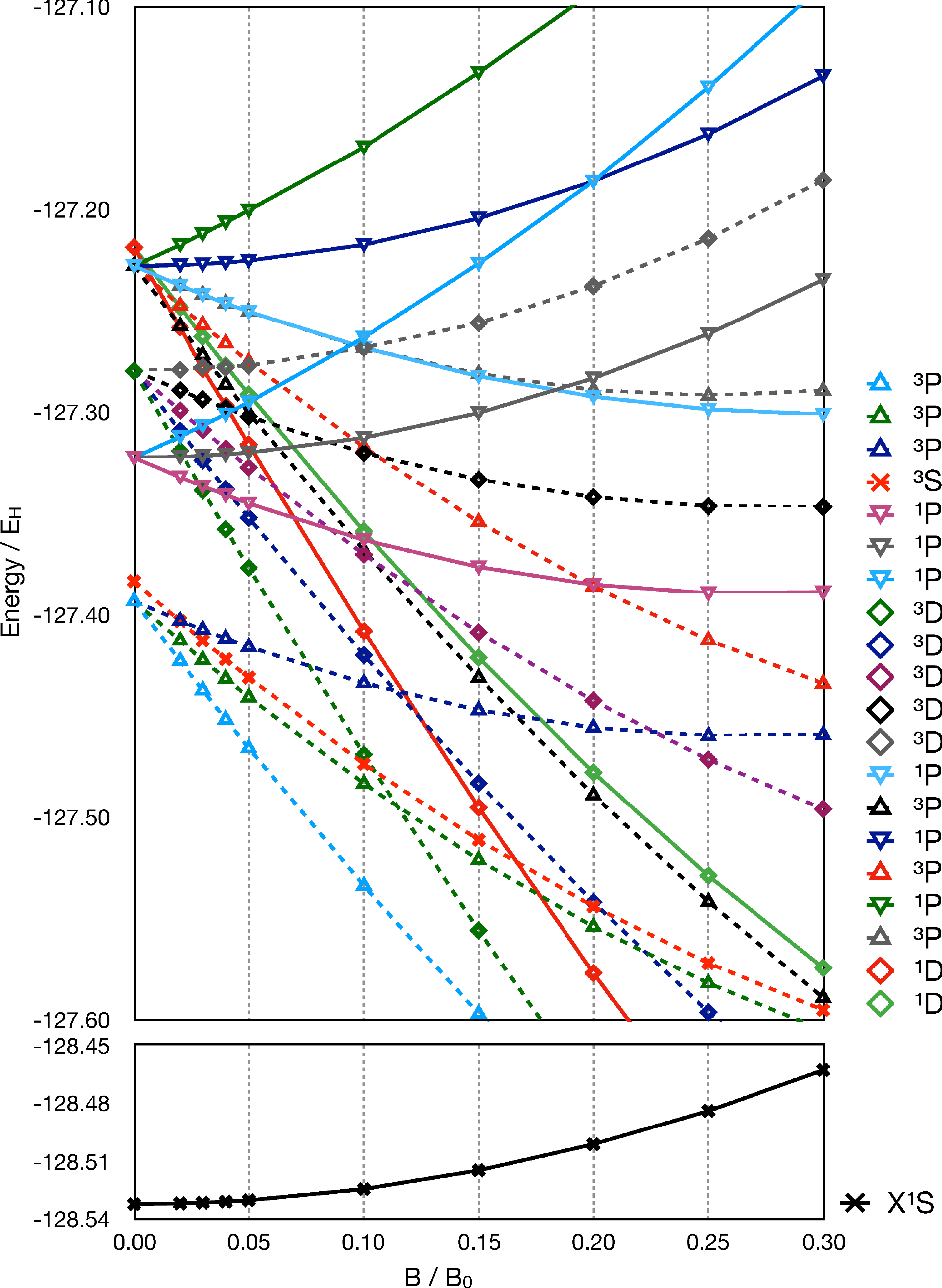}
	\caption{Spectrum of the Ne atom subject to uniform magnetic fields. The ground and excited states are computed using Hartree-Fock (UHF) and RPA respectively with the Lucc-pVTZ basis.}
	\label{fig:neexcbfig1}
\end{figure}

Our first group of examples contains small closed shell atoms He and Ne and the triplet open shell C atom as questions of optimal geometry at each field strength are not a concern in this case.
States are labelled by their zero field term symbols.
The term symbol in the reduced symmetry including the field is mentioned after a slash in some representative cases.
From Fig.~\ref{fig:heexcbfig1}, we note predictable behaviour of the ground and excited states.
The $^1$S states and $^1$P, $m_L$=0 states are diamagnetic, $^1$P states with $m_L= \pm 1$ are orbital paramagnetic and $^3$P states are spin ($m_L=0$) or spin+orbital ($m_L=\pm 1$) paramagnetic.
The competition between the linear and quadratic orbital-Zeeman terms is clearly visible in the $^1$P states with $m_L= \pm 1$ with the energy going through a minimum.
The $^3$P state with $m_L=-1$ becomes the ground state at about $B=0.48$~au.
For $B > 0.57$~au, the zero field ground state becomes a second excited state.
The Ne atom in Fig.~\ref{fig:neexcbfig1} shows a similar predictable behaviour but in this example, we also see the D states.

\begin{figure}
	\centering
	\includegraphics[width=\linewidth]{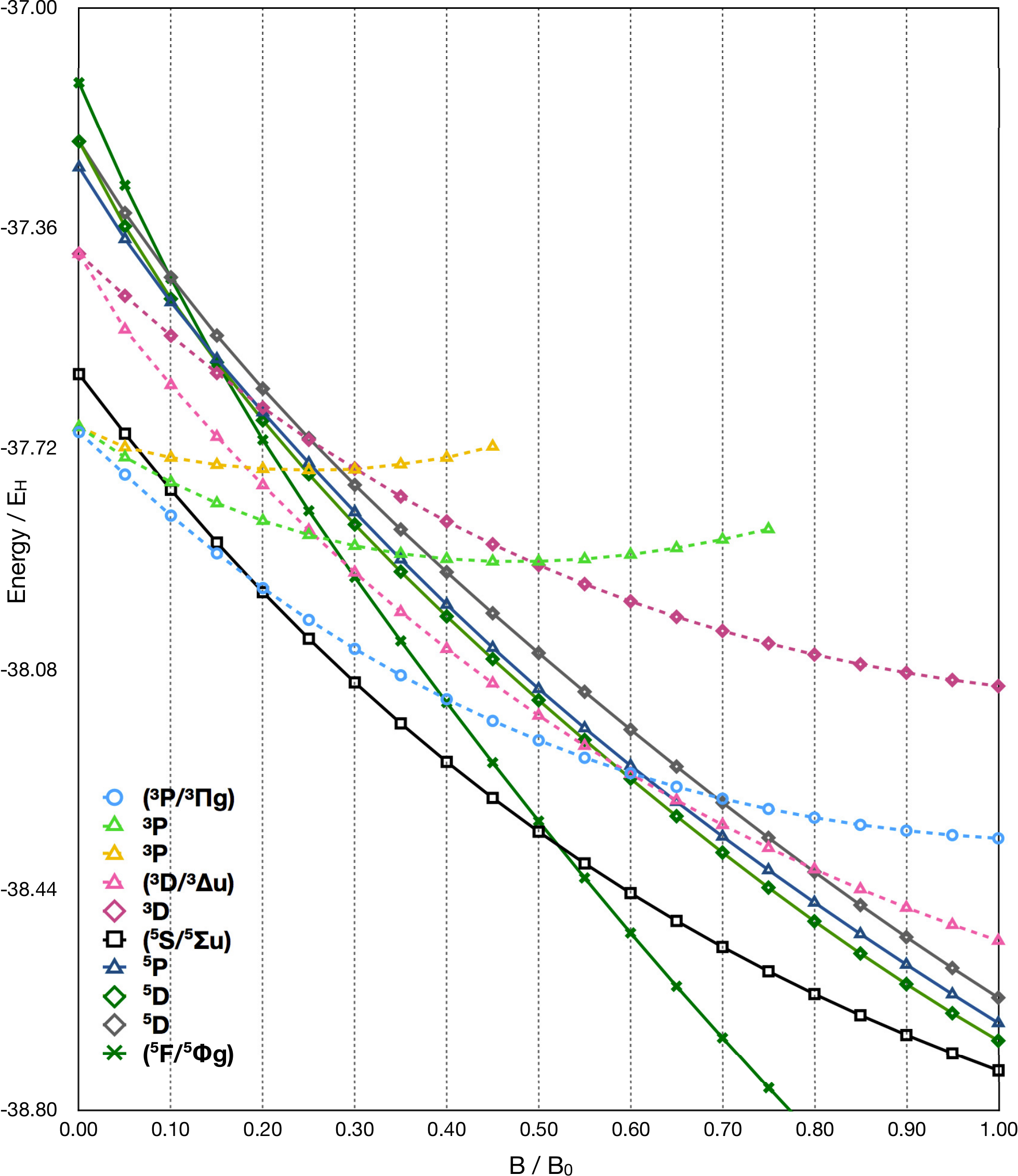}
	\caption{Spectrum of the C atom subject to uniform magnetic fields. The lowest states (at zero field) in each multiplicity are computed with Hartree-Fock (UHF) and the excited states with RPA using the Luaug-cc-pCVQZ basis.}
	\label{fig:cexcbfig1}
\end{figure}

\begin{figure}
	\centering
	\includegraphics[width=\linewidth]{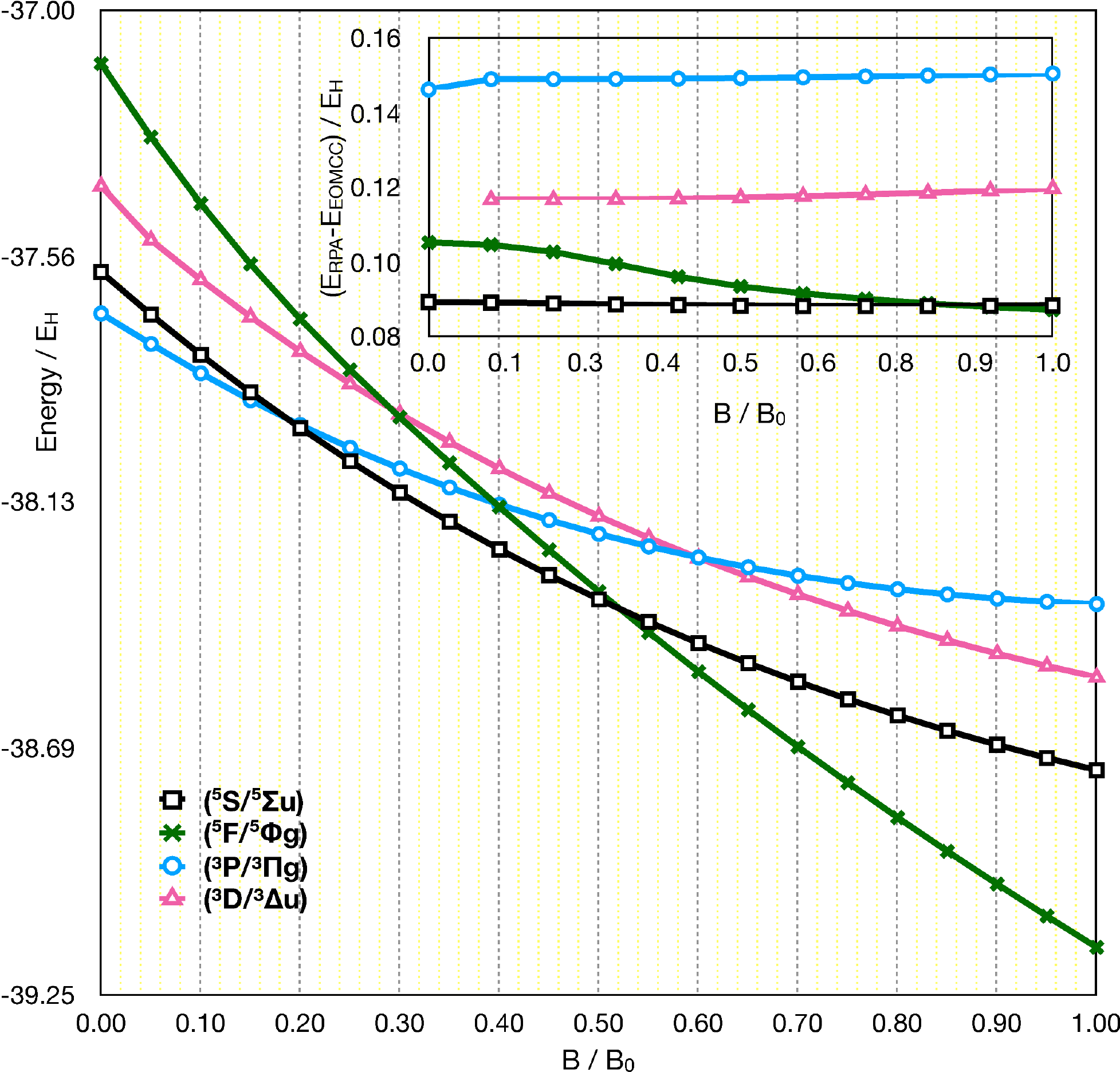}
	\caption{Energy of a few selected states of the C atom subject to uniform magnetic fields. The lowest states (at zero field) in each multiplicity are computed with Hartree-Fock (UHF) and the excited states with RPA using the Luaug-cc-pCVQZ basis. The inset shows the difference of the energies of the excited states computed with RPA and EOM-CCSD.}
	\label{fig:cexcbfigdiffEOM}
\end{figure}

For the computations on the C atom, we have used the $^5$S, $m_S=-2$ UHF function as the reference state for the computation of the pentets and the $^3$P, $m_S=-1$ function for the triplets.
In Fig.~\ref{fig:cexcbfig1}, we see the behaviour of the sets of triplet and pentet states.
The lowest triplet and pentet cross at about $B=0.18$~au, after which the $^5$S, $m_S=-2$ state becomes the ground state.
At about $B=0.52$~au, the $^5$F state overtakes the $^5$S state to become the ground state.
This example is used to benchmark our linear response computations against the recently developed EOM-CC~\cite{Hampe2017}. Hence, we select the states computed by Hampe et al. and plot them in Fig.~\ref{fig:cexcbfigdiffEOM} with the same symbols as in their paper.
The state crossings appear at 0.18~au (42.3 kT) and 0.52~au (122.2 kT) versus 0.31~au (73.6 kT) and 0.51~au (120.6 kT) respectively in EOMCC.
Some state crossings are thus more sensitive to electron correlation than others.
The qualitative behaviour of the states, however, remains the same.
The difference of the state energies between linear response and EOMCC computations are plotted in the inset in Fig.~\ref{fig:cexcbfigdiffEOM} and seem to be reasonably parallel across the range of magnetic fields studied by us. 

\begin{figure}
	\centering
	\includegraphics[width=\linewidth]{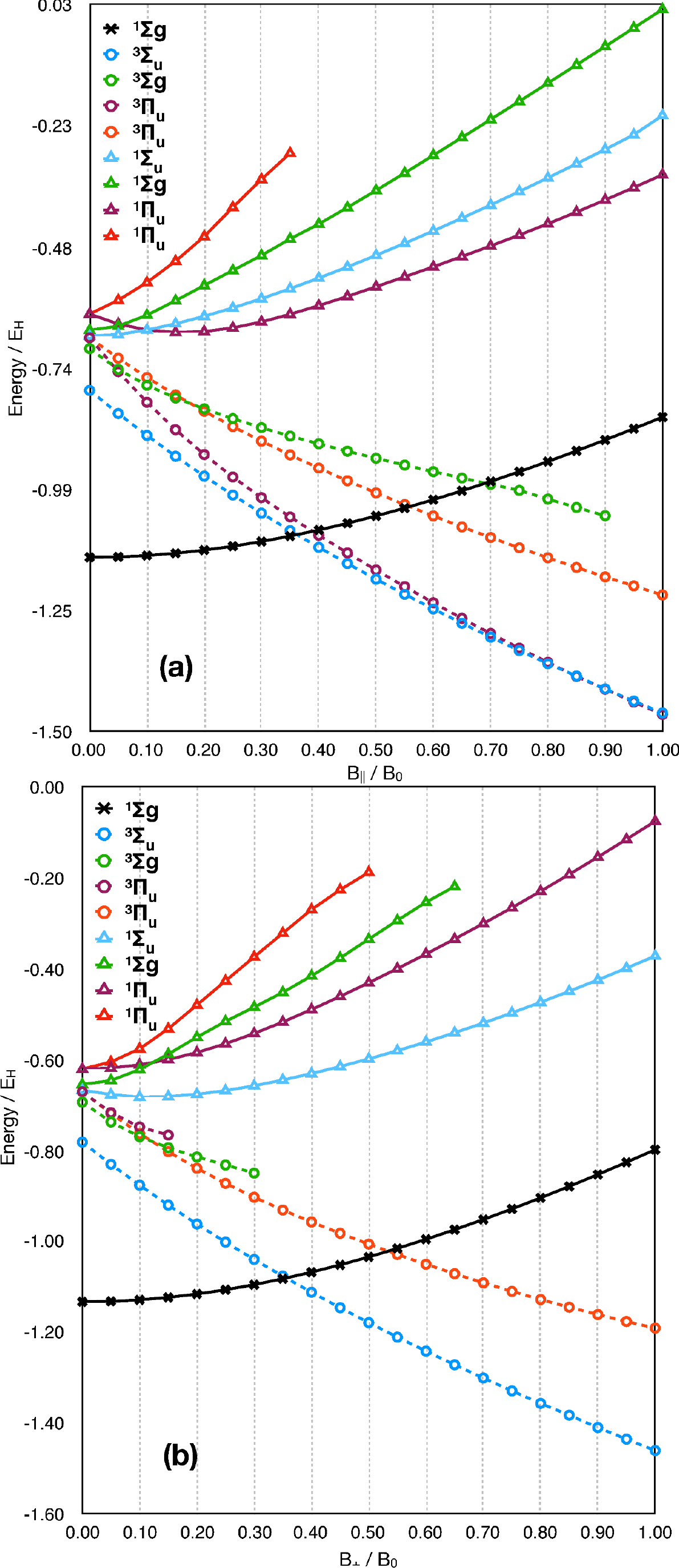}
	\caption[]{Spectrum of H$_2$ subject to uniform magnetic fields (a) parallel and (b) perpendicular to the bond axis. The lowest states (at zero field) are computed with Hartree-Fock (UHF) and the excited states with RPA using the Luaug-cc-pCVQZ basis.}
	\label{fig:h2excb}
\end{figure}

\begin{figure}
	\centering
	\includegraphics[width=\linewidth]{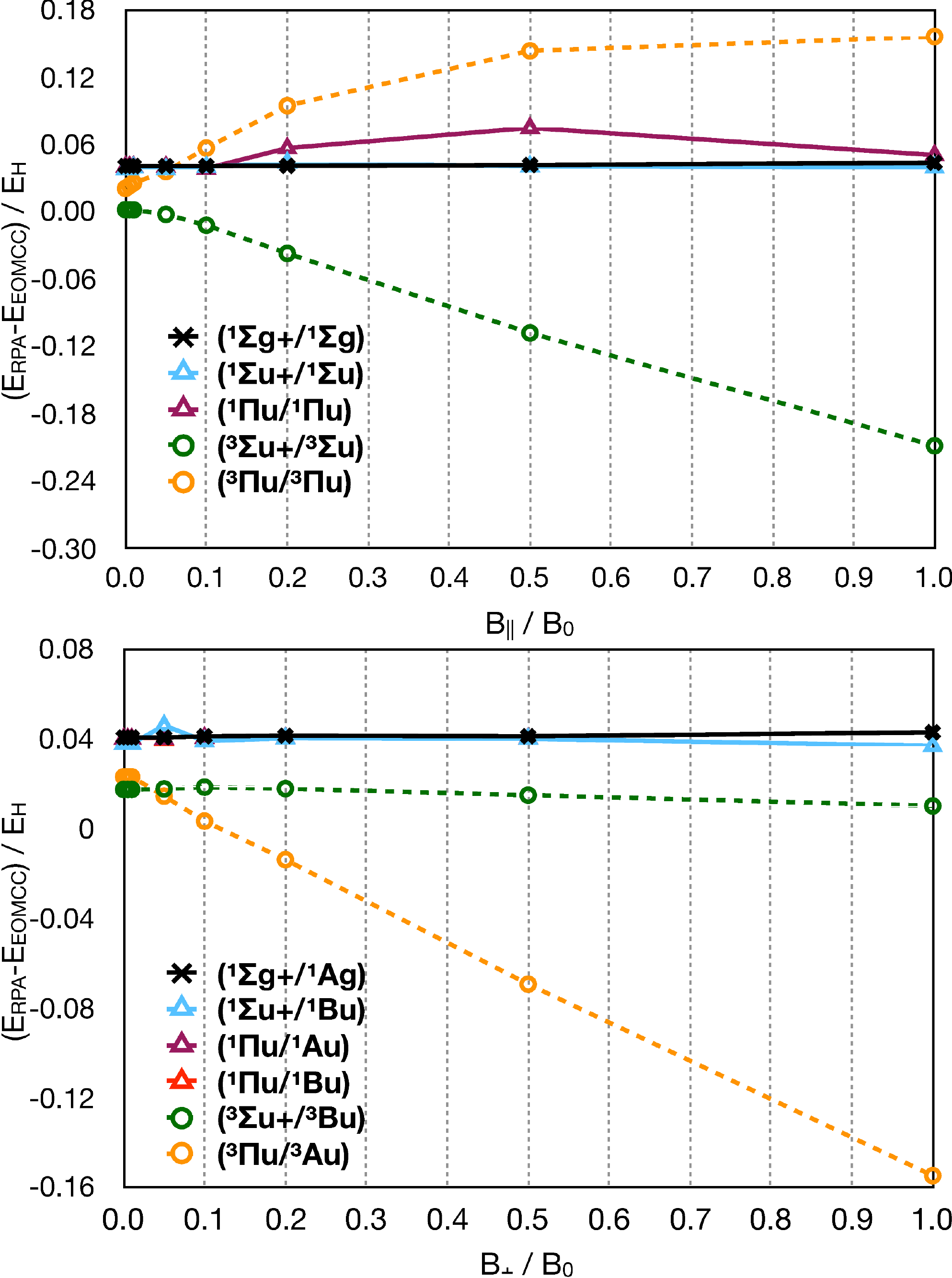}
	\caption[]{Difference of the energies of the ground/excited states computed using UHF/RPA and CCSD/EOM-CCSD of H$_2$ placed in uniform magnetic fields (a) parallel and (b) perpendicular to the bond axis. CCSD optimized geometries at each field are considered.}
	\label{fig:h2excbdiffEOM}
\end{figure}

\begin{figure}
	\centering
	\includegraphics[width=\linewidth]{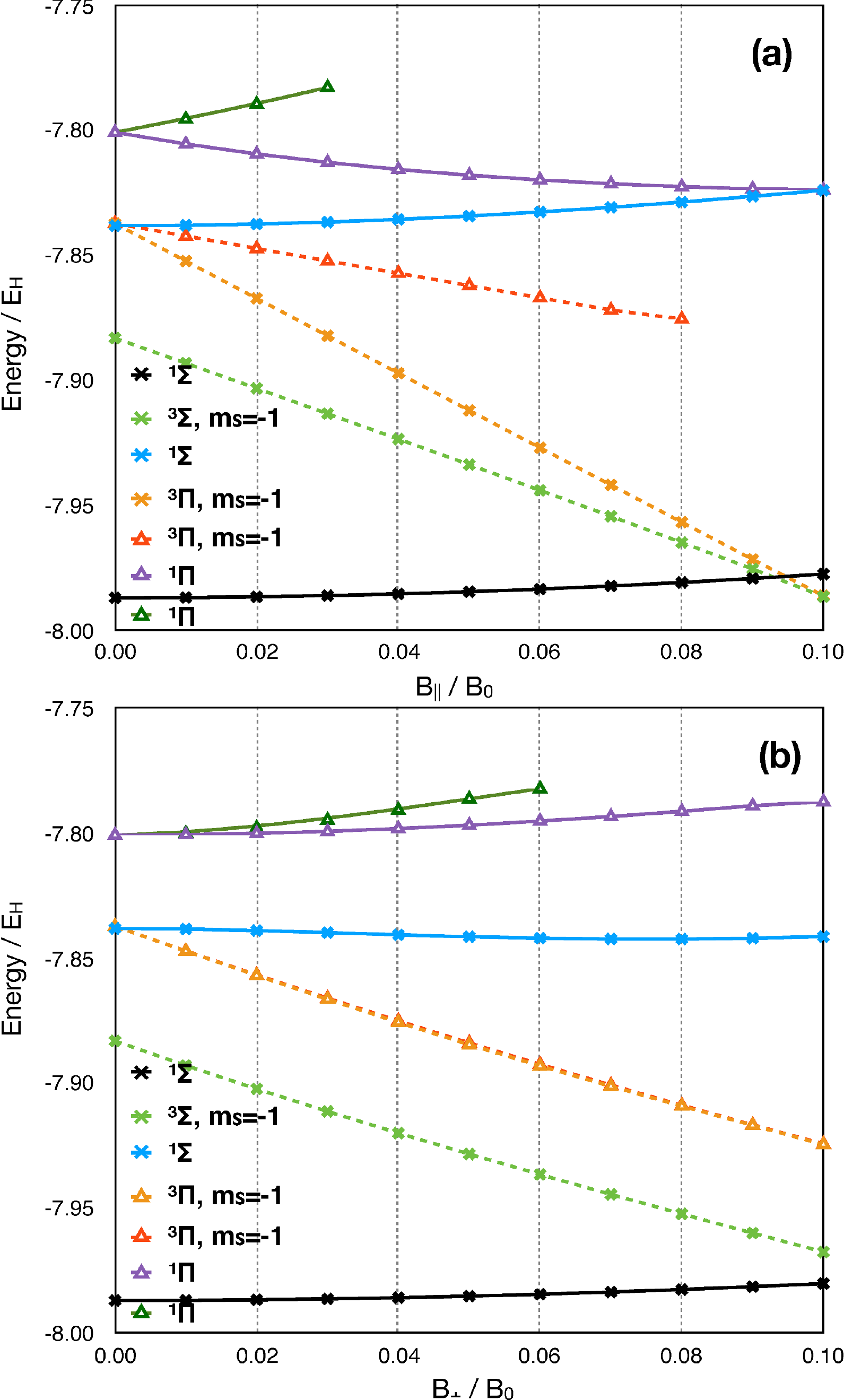}
	\caption[]{Spectrum of LiH subject to uniform magnetic fields (a) parallel and (b) perpendicular to the bond axis. The lowest states (at zero field) are computed with Hartree-Fock (GHF) and the excited states with RPA using the Luaug-cc-pCVQZ basis.}
	\label{fig:liHexcb}
\end{figure}

\begin{figure}
	\centering
	\includegraphics[width=\linewidth]{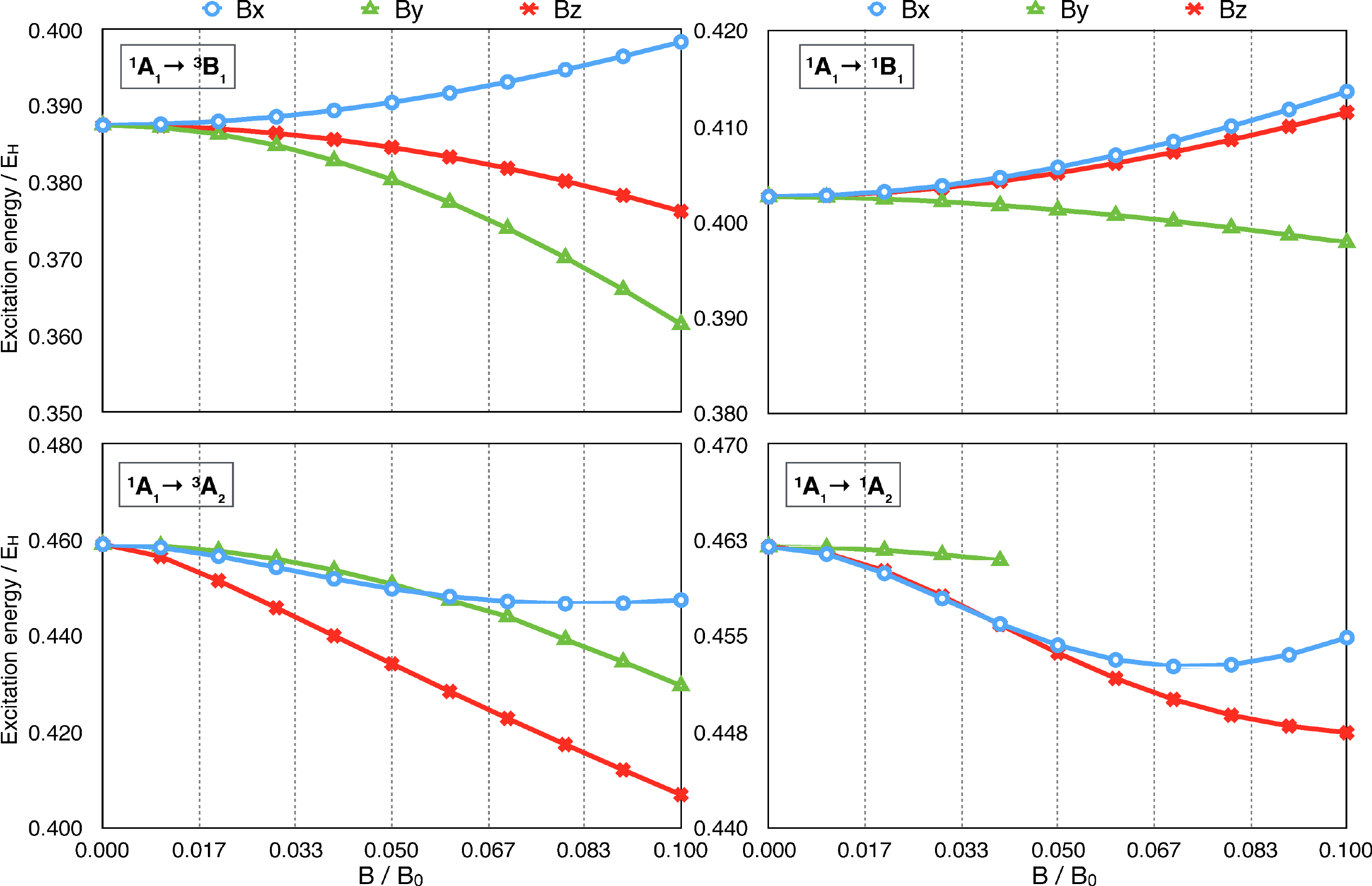}
	\caption{Spectrum of H$_2$O placed in the $yz$-plane is subjected to uniform magnetic fields along the $x$, $y$ and $z$-axis.  The lowest states (at zero field) are computed with GHF and the excited states with RPA using the Luaug-cc-pVDZ basis.}
	\label{fig:h2oexcb}
\end{figure}

Further benchmark studies are carried out on the H$_2$ molecule at a bond length of 1.3984~au, as an example of a small closed shell diamagnetic molecule.
Two orientations of the magnetic field are considered: parallel and perpendicular to the bond axis.
With a parallel field, as shown in Fig.~\ref{fig:h2excb}(a), the ground state switches to the $^3\Sigma_u$ state when $0.36\, \mathrm{au} < B < 0.85\, \text{au}$ and then to $^3\Pi_u$ when $B > 0.85$~au.
The response of $\Pi$ states are found to be stronger than the $\Sigma$ states.
With a perpendicular field, as shown in Fig.~\ref{fig:h2excb}(b), the $^3\Sigma_u$ rapidly comes down with increasing field and becomes the ground state for $B > 0.35$~au.
For the purpose of benchmarking against EOMCC, computations were carried out at a few magnetic field strengths using ground state geometries optimized at CCSD level~\cite{Hampe2017} at each field strength.
The differences of the energies are plotted in Fig.~\ref{fig:h2excbdiffEOM}.
The error curves are more or less parallel except for the $^3\Sigma_u$ state in a parallel field and the $^3\Pi_u$ state in a perpendicular field.
The non-parallelity of these error curves may stem from the correlation energy being strongly dependent on the field strength.

As an example of a small highly polar molecule, we study LiH at a bond length of 3.02356~au and the energy plots are presented in Fig.~\ref{fig:liHexcb}.
A higher sensitivity is noted with crossovers occurring at weaker fields.
For instance, ground and first excited state, $^3\Sigma$, cross around $B_{||} =0.093$~au.

Excitation energies respond differently to fields in different directions.
A representative case is to subject H$_2$O to fields along the
Cartesian axes. The O atom was placed at $(0,0,0.1173)$~bohr and the H
atoms at $(0,\pm0.7572,-0.4692)$~bohr.
In Fig.~\ref{fig:h2oexcb} the variation of the four lowest excitations in H$_2$O placed in the $yz$-plane and having the z-axis as the C$_{2v}$ axis of symmetry, is plotted against $B_x$, $B_y$ and $B_z$.

\begin{figure*}
	\centering
	\includegraphics[width=\linewidth]{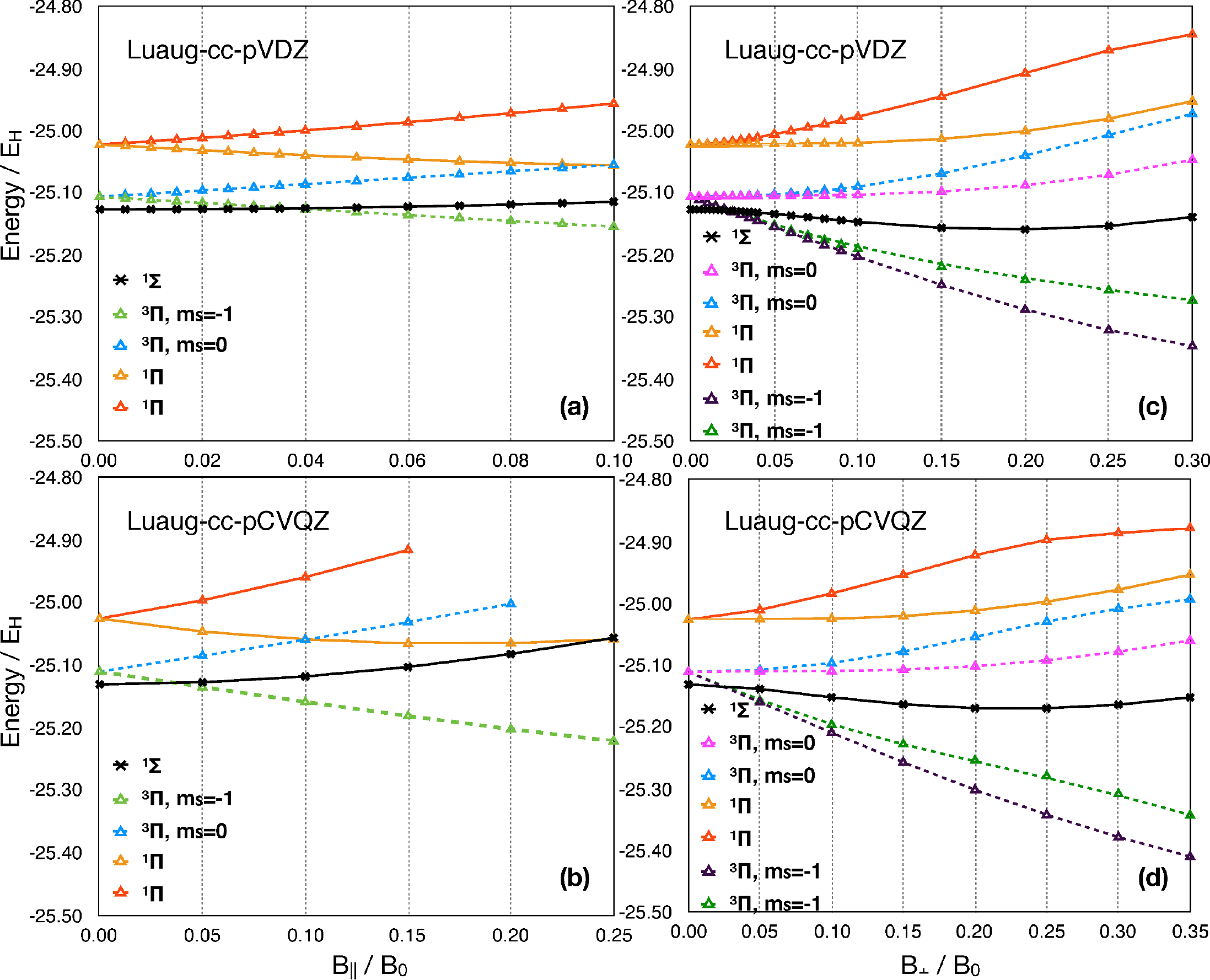}
	\caption{Spectrum of BH in uniform magnetic fields: parallel/perpendicular to the bond computed with TDA in two basis sets, (a)/(c) Luaug-cc-pVDZ and (b)/(d) Luaug-cc-pCVQZ.}
	\label{fig:bhexcb}
\end{figure*}

\begin{figure}
	\centering
	\includegraphics[width=\linewidth]{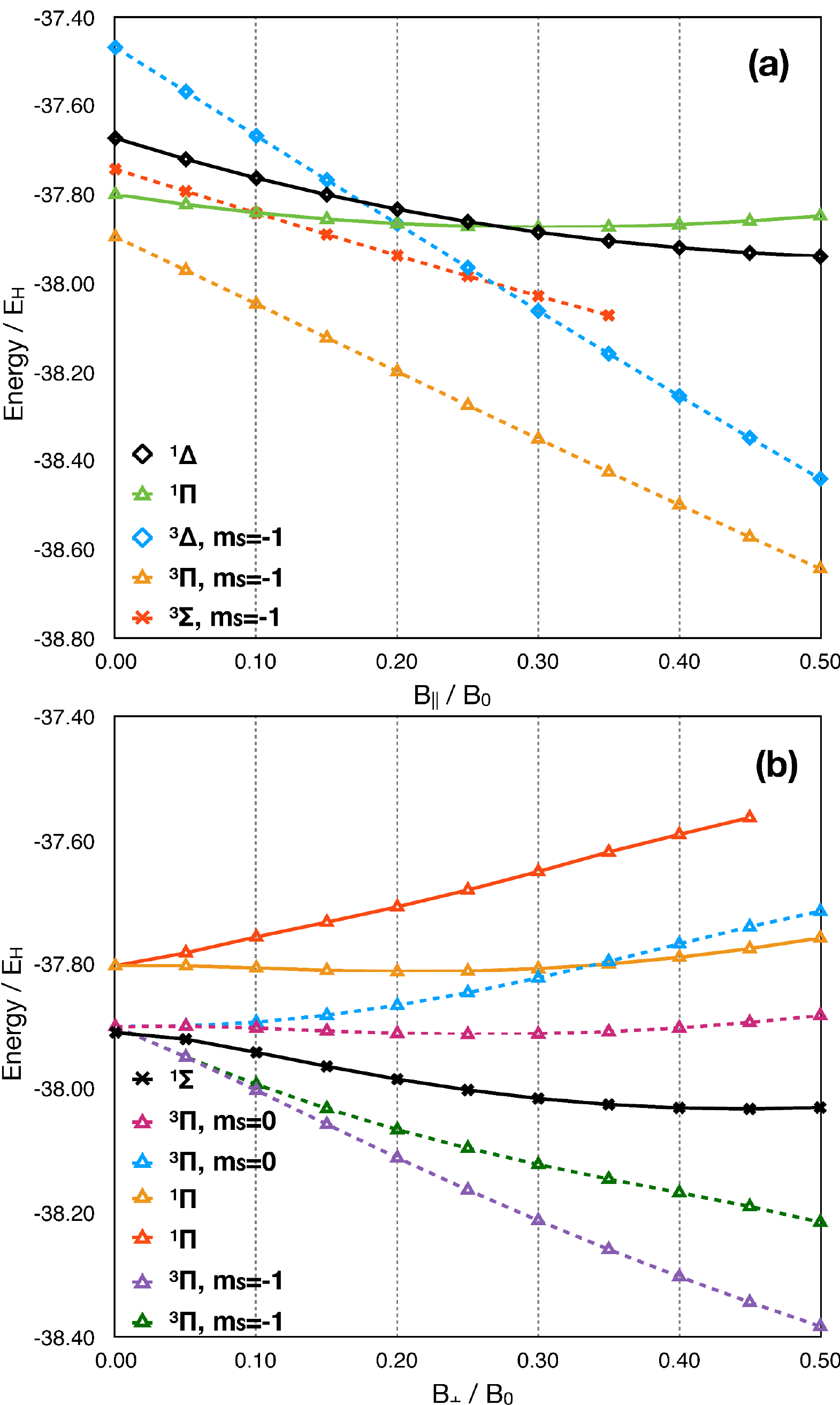}
	\caption{Spectrum of CH$^+$ in uniform magnetic fields (a) parallel and (b) perpendicular to the bond computed with TDA using the Luaug-cc-pCVQZ basis. The $^1\Delta$ GHF state is used as the reference for (a) while the $^1\Sigma$ GHF state is the reference for (b).}
	\label{fig:ch+excb}
\end{figure}

\begin{figure}
	\centering
	\includegraphics[width=\linewidth]{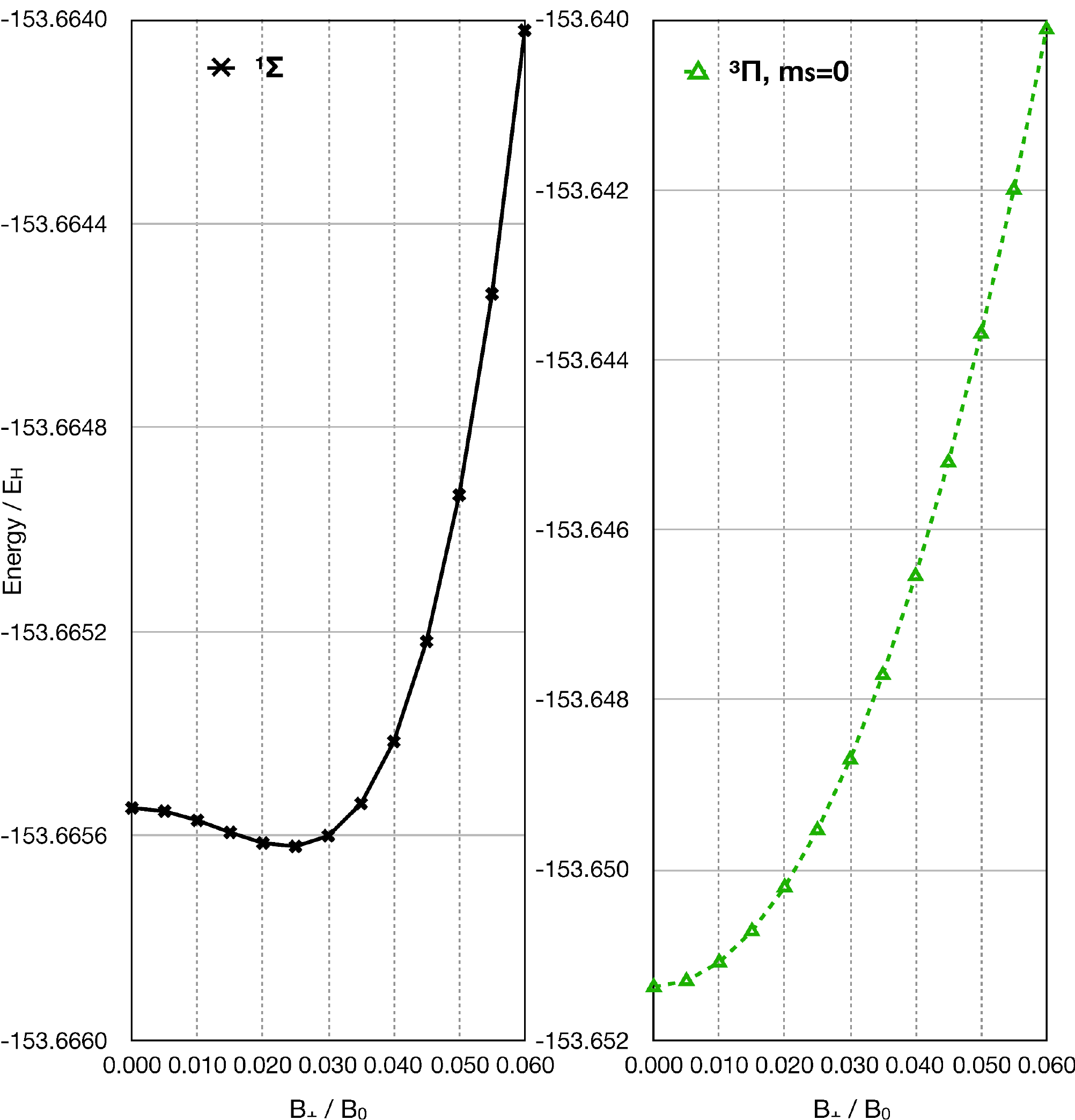}
	\caption{Spectrum of C$_4$H$_4$ placed in a uniform field perpendicular to the plane of the molecule computed with TDA using the Luaug-cc-pVDZ basis.}
	\label{fig:c4h4excb}
\end{figure}

In the next group of examples we consider small closed shell molecules which show paramagnetic behaviour when placed in a perpendicular field.
Due to the inherent triplet instability of the RPA equations, we have adopted the Tamm--Dancoff approximation (TDA)~\cite{Tamm1945,Dancoff1950} for this set of molecules.
BH is our smallest example in this group and we also use it to demonstrate basis set dependence of our computations on account of the sensitivity of it's electronic structure to magnetic fields.
First, we look at the behaviour of BH at a bond length of 2.3342~au in a parallel field. Fig.~\ref{fig:bhexcb}(a) and (b) show the corresponding plots with two basis sets, Luaug-cc-pVDZ and Luaug-cc-pCVQZ respectively.
The plots are qualitatively similar with the ground state crossover from $^1\Sigma$ to $^3\Pi$ occurring at $B=0.035$~au.
In Figs.~\ref{fig:bhexcb}(c) and (d), the ground state shows the characteristic closed shell paramagnetic behaviour at weaker fields and then transitions to diamagnetic behaviour around $B=0.2$~au.
Ground state crossover between $^1\Sigma$ and $^3\Pi$ occurs at $B=0.025$~au with both basis sets.
In parallel fields, the $^3\Pi$, $m_S=0$ states are doubly degenerate while in perpendicular fields, this degeneracy is lifted.
CH$^+$ at a bond length of 2.12122~au shows a very similar spectral behaviour to BH in both a parallel field and a perpendicular field (Fig.~\ref{fig:ch+excb}). 
In the latter situation, the paramagnetic to diamagnetic crossover for the reference state of CH$^+$ happens at $B=0.45$au and a ground-state transition from $^1\Sigma$ to $^3\Pi$ occurs at $B=0.012$~au.
Our next example, in Fig.~\ref{fig:c4h4excb} is rectangular C$_4$H$_4$
with C atoms at $(\pm 1.47588,\pm 1.27462,0)$~bohr and H atoms at $(\pm 2.91248, \pm 2.71626,0)$~bohr. 
It is a paramagnetic closed shell molecule but with a larger cross-sectional area making it more sensitive to perpendicular magnetic fields and, thus, the paramagnetic to diamagnetic crossover for the ground state occurs at a field strength of $B=0.025$~au (left panel) which is an order of magnitude  weaker than BH or CH$^+$.
The $^3\Pi$ excited state is well separated from the ground state in this range of field strengths and flipping of the states would likely occur at fields much larger than the highest field plotted. In a previous study~\cite{Tellgren2009}, C$_4$H$_4$ was found to balance so precisely between (second order) dia- and paramagnetism that the leading order term was quartic in the magnetism field. This can be traced to slightly different geometry compared to the present study.

\subsection{Behaviour of Oscillator Strengths with Changing Magnetic Fields}

\begin{figure}
	\centering
	\includegraphics[width=\linewidth]{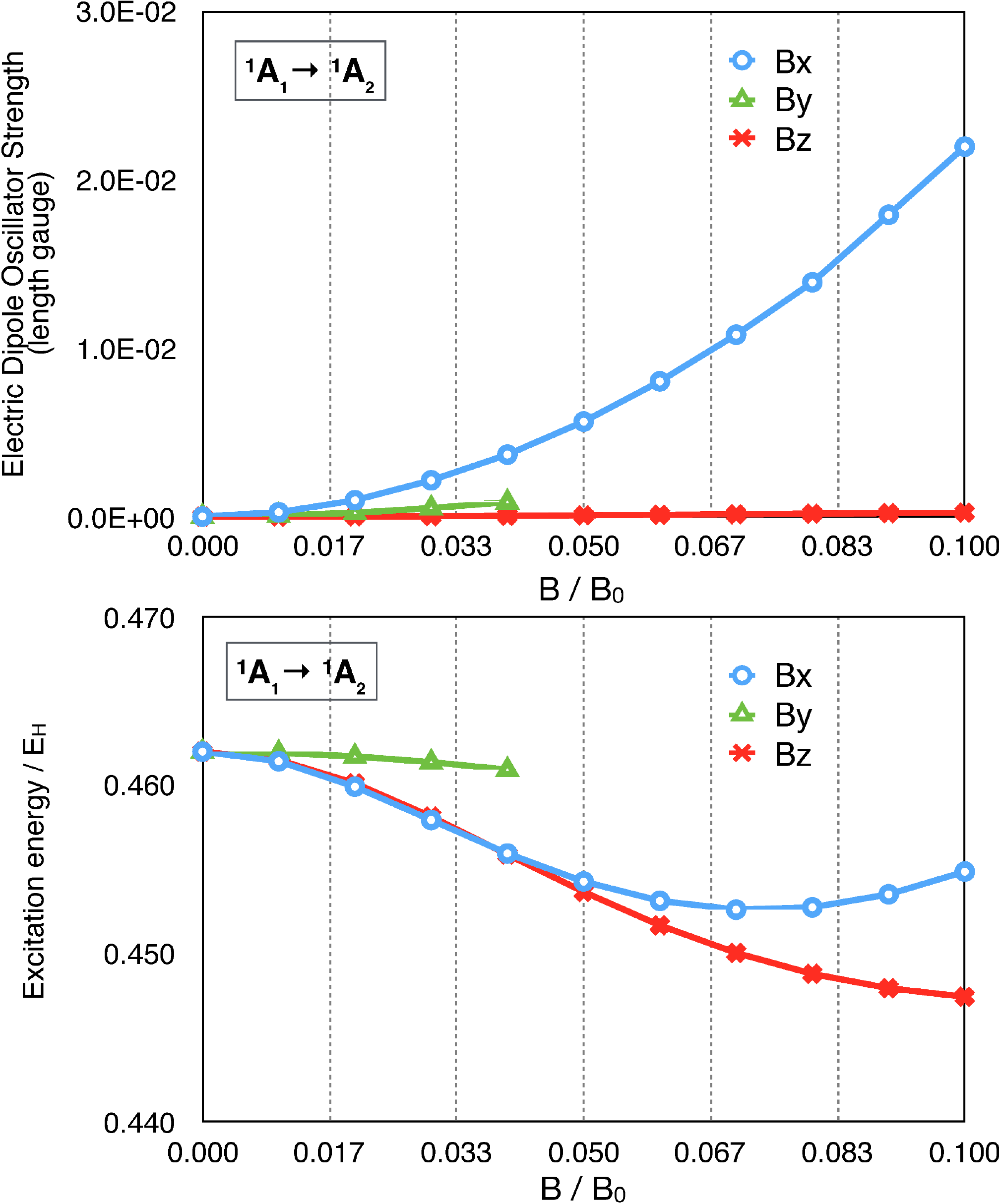}
	\caption{H$_2$O molecule placed in the yz-plane with the z-axis as the C$_{2v}$ axis is subjected to uniform fields along the x, y and z-axis. The spectrum is computed with RPA using the Luaug-cc-pVDZ basis. The spatially forbidden $^1$A$_1 \rightarrow ^1$A$_2$ electric dipole transition becomes allowed when a magnetic field is applied along the $x$ or $y$ direction.}
	\label{fig:h2ooscb}
\end{figure}

\begin{figure}
	\centering
	\includegraphics[width=\linewidth]{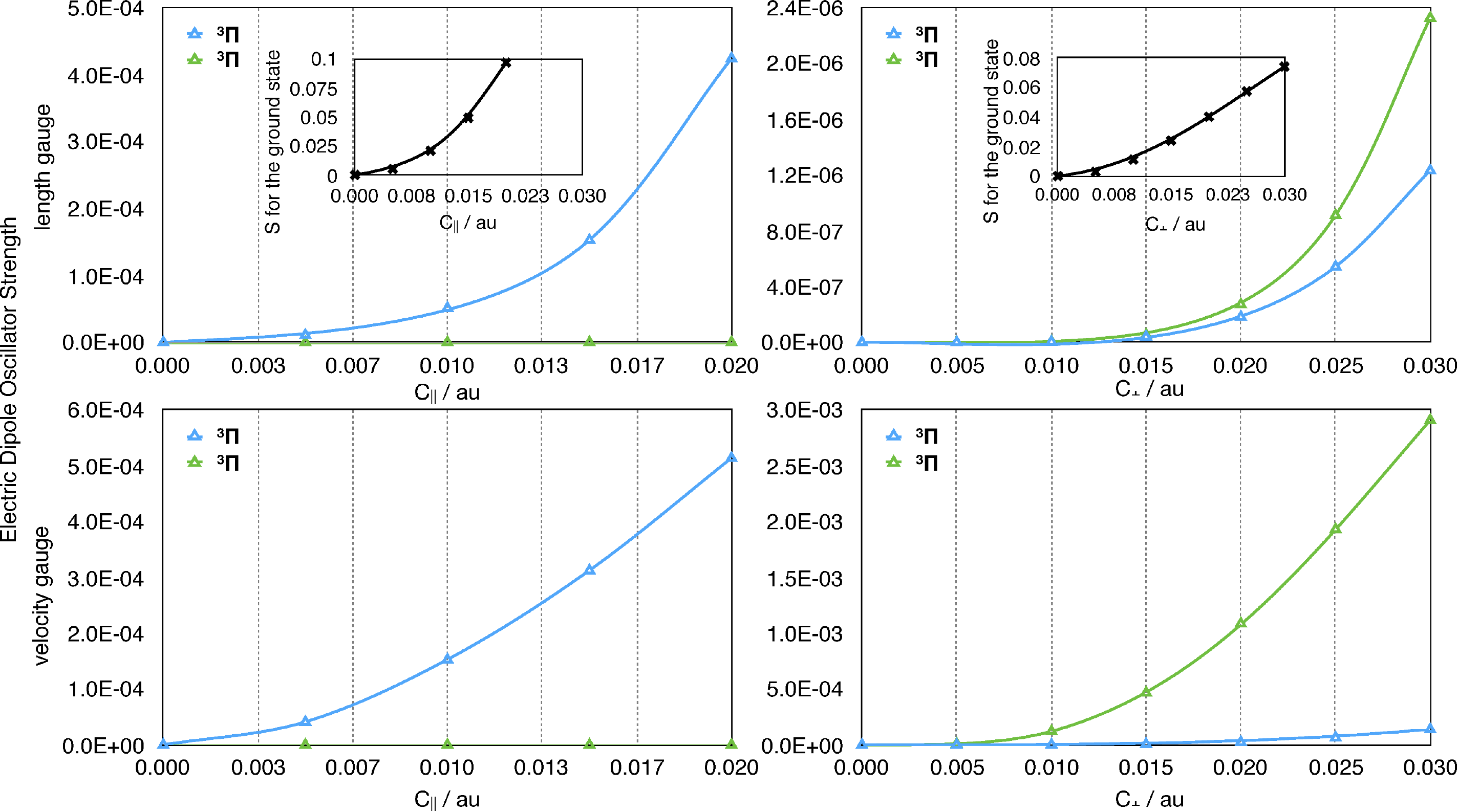}
	\caption{BH molecule placed in a non-uniform field with the curl $\mathbf{C}$ parallel (left panel) or perpendicular (right panel) to the bond axis. Oscillator strengths are computed with TDA using the Luaug-cc-pCVQZ basis. Insets show the spin magnitude $S$ for the reference ground state wave function.}
	\label{fig:bhoscc_ccq}
\end{figure}

\begin{figure}
	\centering
	\includegraphics[width=0.8\linewidth]{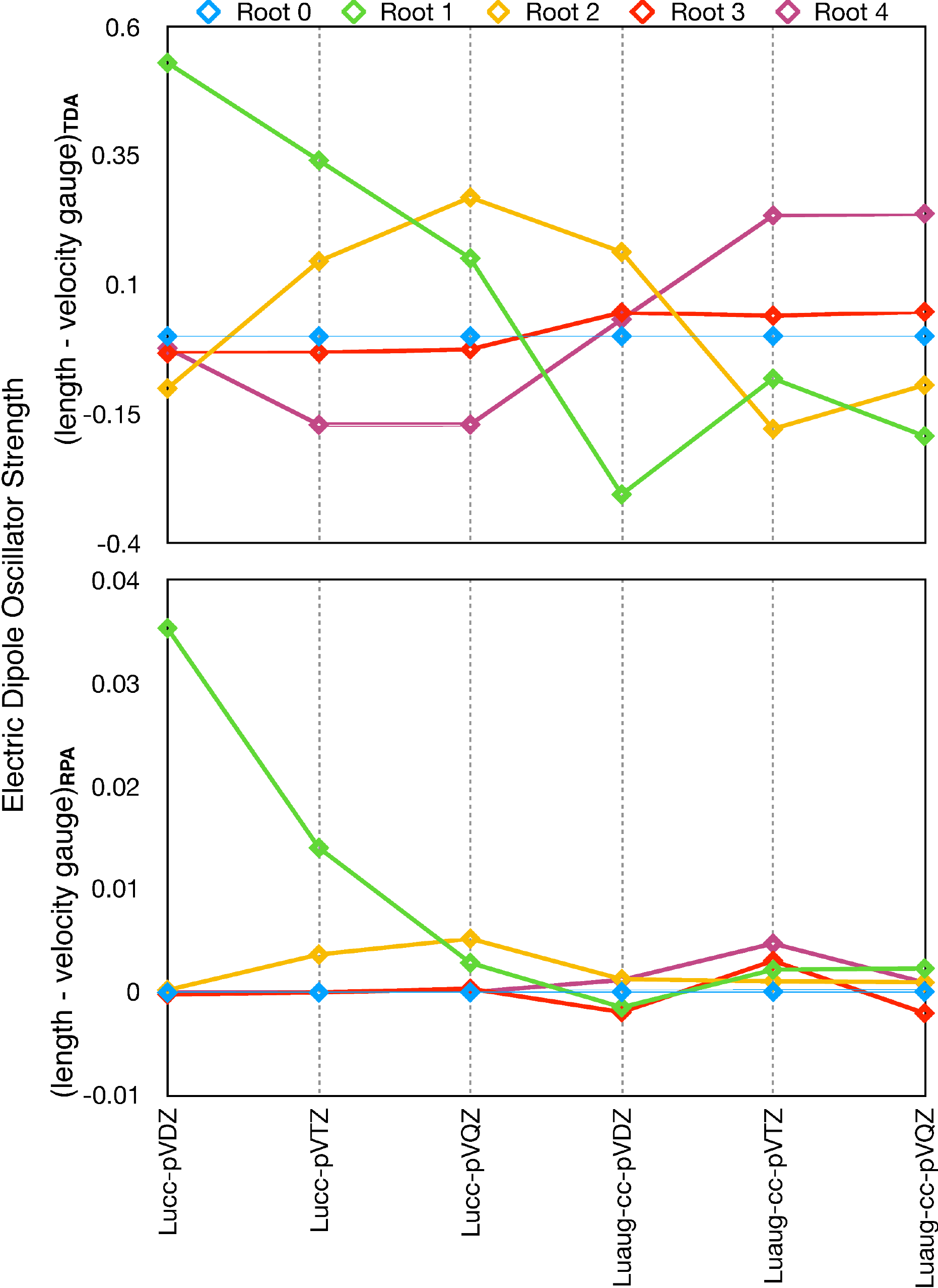}
	\caption{Numerical demonstration of the equivalence of oscillator strengths in the length and velocity gauge in the basis set limit for RPA but not under the Tamm-Dancoff approximation for H$_2$ placed in a non-uniform field with curl $\mathbf{C}=0.03 \hat{\mathbf{e}}_x+0.03\hat{\mathbf{e}}_y+0.03 \hat{\mathbf{e}}_z$.}
	\label{fig:h2osccbasisconv}
\end{figure}

\begin{figure}
	\centering
	\includegraphics[width=0.8\linewidth]{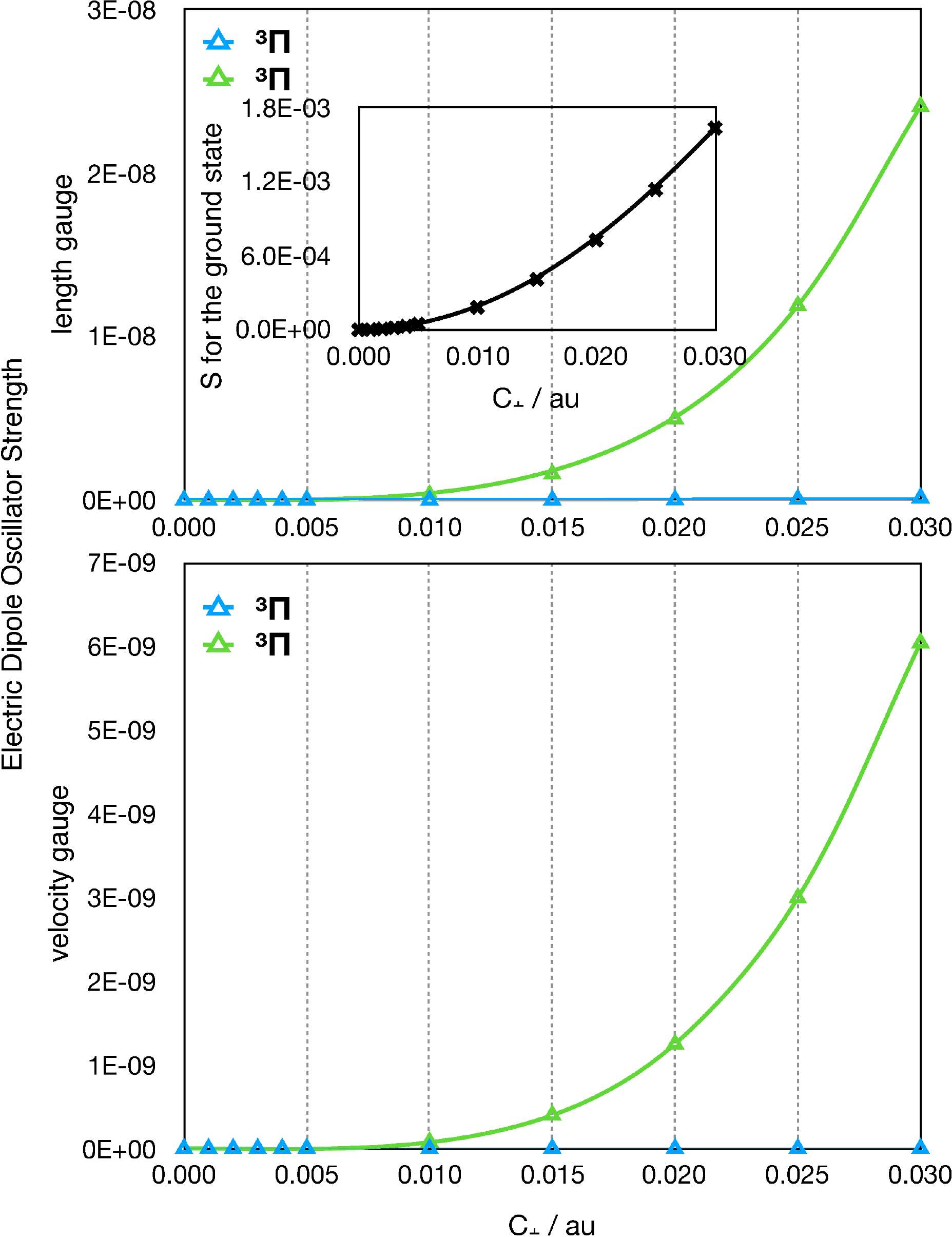}
	\caption{HF molecule placed in a non-uniform field with curl $\mathbf{C}$ perpendicular to the bond axis. Oscillator strengths are computed with RPA using the Luaug-cc-pVDZ basis. The inset shows the spin magnitude $S$ for the reference ground state.}
	\label{fig:hfoscc}
\end{figure}

\begin{figure*}
	\centering
	\includegraphics[width=0.8\textwidth]{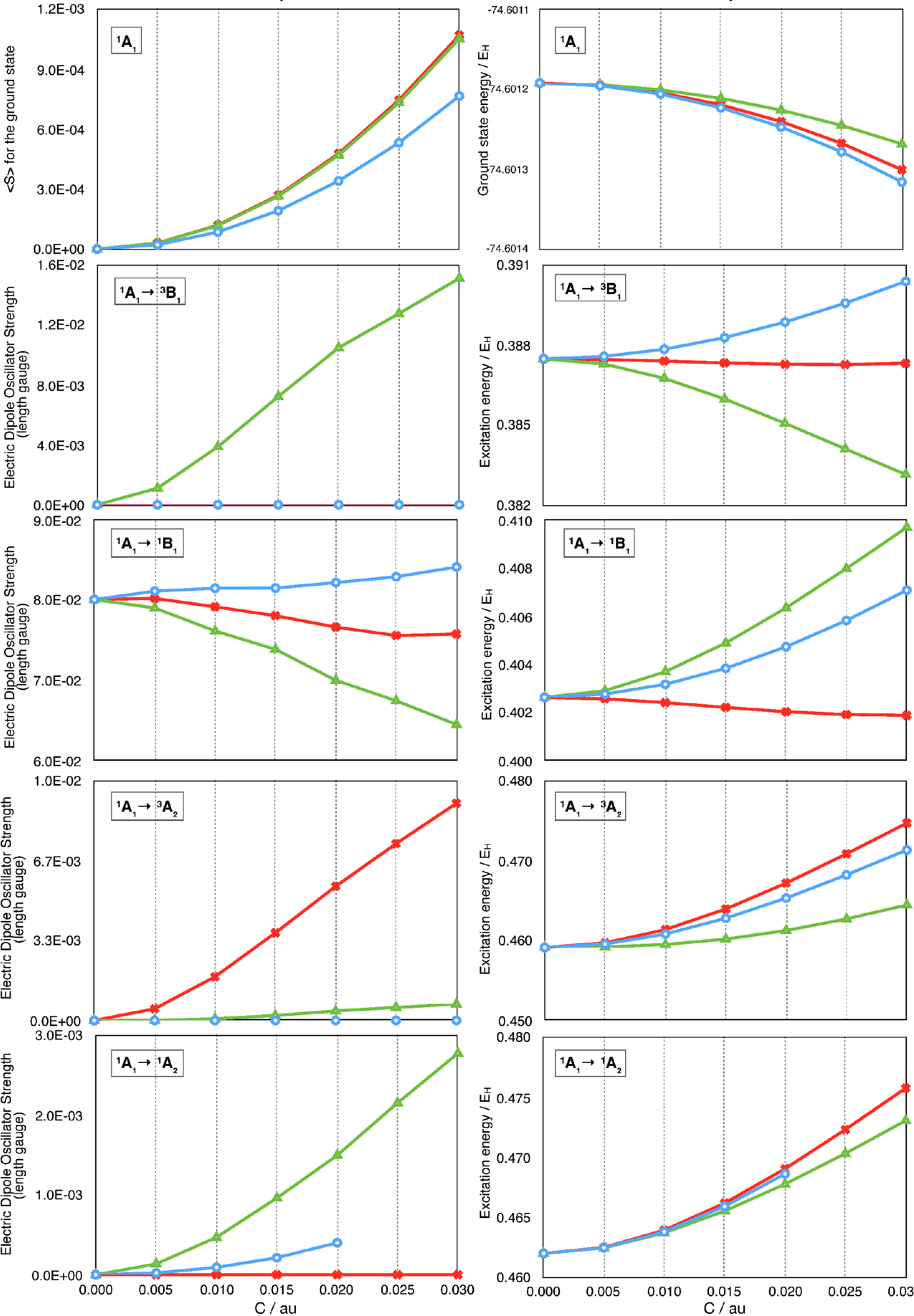}
	\caption{H$_2$O placed in the $yz$-plane and subjected to non-uniform fields where the curl $\mathbf{C}$ is directed along the $x$, $y$ or $z$-axis. Oscillator strengths are computed with RPA using the Luaug-cc-pVDZ basis.}
	\label{fig:h2ooscc}
\end{figure*}

Uniform magnetic fields can break the spatial symmetry of molecules, making spatial-symmetry forbidden transitions, allowed.
A non-uniform magnetic field, such as one with a non-zero curl $\mathbf{C}$ in our case, prevents electronic spins from aligning to a global quantization axis.
A noncollinear spin density is thus generated and $\langle \op{S}^2 \rangle = S(S+1)$ ceases to be a good quantum number.
Normally spin-symmetry forbidden transitions, such as singlet-triplet transitions, thus become allowed.
In this section, we explore both these situations.

Our first example, is H$_2$O placed in a uniform field perpendicular to the plane of the molecule.
The C$_2$ axis and one of the $\sigma_v$ planes of symmetry are thus lost and the $\text{A}_1 \leftrightarrow \text{A}_2$ transitions become electric dipole allowed.
This is demonstrated in Fig.~\ref{fig:h2ooscb}.

The second set of examples, involves non-uniform fields where we focus on the lowest singlet-triplet transitions in a variety of small molecules.
The most sensitive singlet-triplet transition appears to be those into $\Pi$ orbitals, such as $\Sigma \rightarrow \Pi$ and $\text{n} \rightarrow \Pi$* transitions.
BH shows a triplet instability in the response computations and the full RPA computation collapses. We have thus adopted the Tamm-Dancoff approximation (TDA) in this case.
The electric dipole oscillator strengths for the $^1\Sigma \rightarrow {}^3\Pi$ transitions are presented in Fig.~\ref{fig:bhoscc_ccq}. 
The increase in oscillator strength follows the trend of the deviation of the spin magnitude $S$, calculated by inverting $\langle \op{S}^2 \rangle = S(S+1)$, from zero (shown as inset) in both the length and the velocity gauge.
However, due to the inequivalence of the two gauges in the TDA which is basically a singles configuration interaction (CI), the values are widely different.
In fact, when the curl of the external magnetic field is perpendicular to the bond axis (left panel of Fig.~\ref{fig:bhoscc_ccq}), they differ by several orders of magnitude.

Equivalence of the length and velocity gauge in oscillator strength computations using RPA is guaranteed in the basis set limit when the orbitals are real~\cite{Jorgensen1975}.
With complex orbitals, as in our case, no proof has been put forth to the best of our knowledge.
In Fig.~\ref{fig:h2osccbasisconv}, we numerically demonstrate that oscillator strengths in the length and velocity gauge do indeed converge for the RPA (bottom panel) even with complex orbitals but fail to do so in the TDA (top panel).
A highly inhomogeneous magnetic field with $\mathbf{C}=0.03 \hat{\mathbf{e}}_x+0.03\hat{\mathbf{e}}_y+0.03 \hat{\mathbf{e}}_z$ has been used.
The basis set convergence in the presence of magnetic fields is found to be slower than in the zero-field case.

The $\Pi\rightarrow\Sigma^*$ transitions in the HF molecule at a bond length of 1.7325~au are also sensitive to the breaking of spin symmetry.
In Fig.~\ref{fig:hfoscc}, the oscillator strength is seen to rise rapidly with the increasing spin magnitude $S$ that is generated by a curl $\mathbf{C}$ perpendicular to the bond axis.
The excitations in H$_2$O show diverse behavior in response to different field inhomogeneities, as seen in Fig.~\ref{fig:h2ooscc}.
While the increase in $S$ is similar when the curl $\mathbf{C}$ is directed along either of the Cartesian $x$, $y$ and $z$ directions, the nature of the excited state determines how it is affected by the various orientations of $\mathbf{C}$. 

\section{Conclusion}

In this paper, we report an implementation of the random phase approximation theory using complex orbitals to compute the electronic spectra of molecules placed in a strong external magnetic field which may be uniform or non-uniform. Two-component orbitals are required in the latter case. We compute the electronic excitation energies of small molecules including those of astrochemical importance and benchmark RPA against EOMCC wherever applicable. We find the qualitative behaviour of RPA to be similar to EOMCC with largely parallel error curves over a wide range of uniform magnetic fields. Change of the ground states of molecules from the closed shell singlet to progressively states of higher spin multiplicity is expected and is generally observed. Excited states are found to be more sensitive to changes in magnetic fields. Polar molecules like LiH show larger responses at weaker fields.

Our study of oscillator strengths for the lowest singlet-triplet transition of closed shell molecules, indicates that the growth from zero value with increasing non-uniformity of the magnetic field roughly follows the deviation of $S$ from zero. However, for less symmetric excited states, the exact behaviour depends on the particular excitation involved. The equivalence of the length and velocity gauge at the basis set limit in RPA computations with complex orbitals has also been numerically demonstrated.

\section*{Acknowledgments}

This work was supported by the Research Council of Norway through Grant No.~240674 and CoE Hylleraas Centre for Molecular Sciences Grant No.~262695, and the European  Union's Horizon 2020 research and innovation programme under the Marie Sk{\l}odowska-Curie grant agreement No.~745336. This work has also received support from the Norwegian Supercomputing Program (NOTUR) through a grant of computer time (Grant No.~NN4654K).

\newpage
\clearpage

\end{document}